\documentclass[useAMS,usenatbib]{mnras}

\usepackage[dvips]{graphics}
\usepackage{graphicx}
\usepackage{amsmath}
\usepackage{epsfig}
\usepackage[latin1]{inputenc}
\usepackage{hyperref}

\usepackage[T1]{fontenc}
\usepackage{ae,aecompl,url}
\usepackage{multicol,placeins}
\usepackage{cuted,mathtools}

\title[]{Migration in the shearing sheet and estimates for young open cluster migration}

\author[Quillen et al.]
{
Alice C. Quillen$^{1}$, 
Eric Nolting$^{1}$,
Ivan Minchev$^2$,  
\newauthor
Gayandhi De Silva$^{3,4}$ \&
Cristina Chiappini$^2$ 
\newauthor
\\
$^1$Department of Physics and Astronomy, University of Rochester, Rochester, NY 14627 USA \\
$^2$Leibniz-Institut f\"ur Astrophysik Potsdam (AIP), An der Sternwarte 16, D-14482, Potsdam, Germany\\
$^3$Australian Astronomical Observatory, 105 Delhi Rd, North Ryde, NSW 2113, Australia\\
$^4$Sydney Institute for Astronomy, School of Physics, A28, The University of Sydney, Sydney NSW 2006 Australia}

\begin{document}
\maketitle

\begin{abstract}

Using tracer particles embedded in self-gravitating shearing sheet N-body simulations,  we investigate the distance in guiding centre radius that stars  or star clusters can migrate in a few orbital periods.   The standard deviations of guiding centre distributions and maximum migration distances depend on the Toomre or critical wavelength and the contrast in mass surface density caused by spiral structure.  Comparison between our simulations and estimated guiding radii for a few young super-solar metallicity open clusters, including NGC 6583, suggests that the contrast in mass surface density in the solar neighbourhood has standard deviation (in the surface density distribution) divided by mean of about 1/4 and larger than measured using COBE data by Drimmel and Spergel.  Our estimate is consistent with a standard deviation of $\sim$0.07 dex in the metallicities measured from high-quality spectroscopic data for 38 young open clusters ($<$1 Gyr) with mean galactocentric radius 7-9\ kpc.


\end{abstract}



\section{Introduction}

In most cases, the stellar surface abundances reflect the composition of the interstellar medium at the time of their birth; so stars can be viewed as fossil records of galaxy evolution.
Open clusters abundances probe the chemical evolution of the Galactic thin disc 
(e.g., \citealt{janes79,friel95,desilva06,friel10,pancino10,yong12,magrini15,jacobson16,netopil16,anders17,cantatgaudin16,magrini17,casamiquela17,casamiquela17b}).  
From cluster ages,  their abundances and galactocentric radii,  the galactocentric radial metallicity gradient,
the metallicity  scatter and the time evolution of these quantities 
(e.g., \citealt{loebman16,jacobson16,netopil16,anders17})
can be compared to  chemical evolution models (e.g., \citealt{chiappini01,minchev13,minchev14})
 so as to improve understanding of how the Galaxy assembled  and evolved.   
 
Nearby and young early B stars in the solar neighborhood are chemically homogeneous, suggesting
that the local interstellar medium,  from which stars form, is quite homogeneous chemically \citep{przybilla08,nieva12}.
From the B star homogeneity we infer that
mixing in the interstellar medium is efficient and thorough (e.g.,  \citealt{feng14}). 
The nearby B stars have iron abundance [Fe/H] $= 0.02 \pm 0.04$,
equivalent to, within the estimated uncertainties,
the iron abundance of the Sun (using values from Table 9 by \citealt{nieva12}
and the Solar iron abundance by \citealt{asplund09}).
Slow variations in stellar or gas iron abundance, [Fe/H], are often described solely 
with a radial metallicity gradient where the gradient depends on the derivative with
respect to galactocentric radius.
Migration of stars or clusters from their birth radius \citep{wielen77,wielen96,sellwood02,jilkova12} 
broadens  local age and metallicity distributions 
\citep{roskar08,schonrich09,stanghellini10,loebman11,brunetti11,roskar12,minchev13,haywood13,minchev14,loebman16}. 
We use the term {\it radial migration} to refer to a dynamical process that slowly varies the mean orbital
galactocentric radius of a star or cluster that is part of the rotating disc in a disc galaxy.
Both stars and star clusters can radially migrate (e.g., \citealt{jilkova12}).
The expected chemical enrichment in the last 4 Gyrs is around or below 0.1 dex 
 for alpha-elements, and around 0.15--0.2 for iron peak elements 
\citep{chiappini03,asplund09}. The presence of super metal rich stars (with [Fe/H] $>$ 0.25) within the solar neighbourhood \citep{soubiran99} cannot be explained from a baseline chemical evolution model, 
without requiring radial migration  \citep{chiappini09,casagrande11,minchev13}.

Open clusters are a setting where observations can be combined to give
both age and metallicity measurements.
A number of open clusters are so metal rich that they have super-solar metallicities, 
including NGC 6253 \citep{carretta07,maderak15,netopil16}, NGC 6791 \citep{carretta07,peterson98,casamiquela17}, 
 NGC 6583 \citep{magrini10} and NGC 6067 \citep{alonsosantiago17}.
Older metal rich open clusters such as 
NGC 6253 (age 3.3 Gyr, \citealt{maderak15}) or NGC 6791 (age 8 Gyr, \citealt{anthonytwarog10}) 
could have been
born from initially more metal rich gas located close to the Galactic bulge, and then migrated outward
(e.g., \citealt{jilkova12}).
Alternatively parent molecular clouds could have been locally enriched by nearby
supernovae prior to cluster formation (see discussions by \citealt{maderak15,magrini15}).
The two scenarios might be told apart from patterns in  $\alpha$ process  and iron peak element
abundances.
The recent study by \citet{magrini17} finds agreement between age, radius
and abundance distributions of open clusters and the predictions of
chemical evolution models that are based on N-body numerical simulations of a Milky Way-like galaxy
 that exhibit radial migration  \citep{minchev13,minchev14}. 

Young super-solar metallicity open clusters cannot be too far from their birth galactocentric
radii as the process
of radial migration has less time to operate.
Using the values recently tabulated by \citet{bland16}, the rotation period
at the galactocentric radius of the Sun ($R_\odot \approx 8.2$ kpc) is approximately 
210 Myr (using angular rotation rate $\Omega_0 = 30 {\rm km~s}^{-1}{\rm kpc}^{-1}$).
An open cluster that is 1 Gyr old, such as NGC 6583 \citep{carraro05}, 
 would rotate about the Galaxy only approximately
5 rotation periods during its lifetime (using the rotation period near the Sun).   
The Hyades and Praesepe (NGC 2632)
clusters with age approximately 700 Myr \citep{cummings17}  have 
iron abundance [Fe/H] $\approx 0.15$ \citep{cummings17}  and
ages corresponding to three rotation periods.
If these clusters were born at smaller galactocentric radii,  the difference
between their estimated birth radius and current mean orbital radius
must constrain the extent of radial migration possible in a few rotation
periods (e.g., see discussion by \citealt{bland10}).
 
We focus here on whether and how star or star-cluster radial migration could occur in a Gyr.
If a transient spiral pattern grows and decays on a timescale comparable to 
one half the oscillation period within a horseshoe orbit of the corotation region of a spiral wave, 
a star or star cluster can be moved from one side to the other side of the 
corotation resonance.  The star or star cluster is left on the other side of resonance if 
the spiral pattern vanishes before pulling it back \citep{sellwood02}.
This mechanism is often called {\it radial migration} and when caused by
 stochastic growth and disappearance of transient spiral waves it is sometimes called `churning' \citep{sellwood02,roskar08,schonrich09}.    Additional mechanisms 
may induce radial migration,  such as resonant coupling
with bar and spiral arms \citep{minchev10,brunetti11} and interference between
spiral patterns \citep{quillen11,comparetta12}. 

The radial maximal migration rate is expected to depend on the surface density and amplitude
of spiral structure \citep{sellwood02,schonrich09,daniel15}.  
The 3D stellar structure of the Milky Way based on COBE/DIRBE data \citep{drimmel01}
found an on-off surface density contrast in the stellar component of the strongest spiral arm (namely
the Crux-Scutum arm at a galactic longitude of about $310^\circ$)
$(\Sigma_{max} - \Sigma_{min})/ \Sigma_{min} \sim 0.32$
and this is below that expected for spiral galaxies similar to the Milky way (of order 1,  e.g., \citealt{ma02}).
The Glimpse survey observations confirmed the COBE/DIRBE spiral tangent arm detections 
\citep{benjamin05} but have not yet updated an estimate for the stellar spiral arm surface density contrast.
We can consider the amplitude in surface density of spiral structure between  $R_\odot$  and the bar
end at about 4 kpc as poorly constrained.   
We ask here: Is the roughly measured amplitude in spiral structure in the Galaxy large enough
to achieve migration rates necessary to account for young super-solar metallicity open clusters? 
The answer to this question would help us differentiate between local enrichment
and migration processes, and connect the age, metallicity and orbit distributions of open clusters 
 to migration models.

A number of studies have measured radial migration rates from numerical simulations
(e.g., \citealt{minchev11,brunetti11,loebman11,roskar12,comparetta12,grand12,minchev12,
minchev13,minchev14,loebman16,martinez17}).
A difficulty of using N-body simulations of an entire galaxy is that
underlying parameters such as spiral amplitude are difficult to adjust.
Instead of an N-body simulation that simulates an entire disc,
we focus on a small patch of the disc using the shearing sheet approximation 
\citep{julian66,toomre91,rein12};
(see Figure \ref{fig:shearsheet} and our appendix).  
The shearing sheet is a model dynamical system that can be used to study the  dynamics 
of astrophysical discs.    A self-gravitating shearing sheet exhibits 
spiral instability \citep{julian66,toomre81,toomre91}.
The advantage of focusing on a small patch is that the simulation is independent
of galactic radius and depends on 
only a few parameters.  We aim to 
 adjust the amplitude of spiral structure to
 probe how far stars or star clusters can migrate in few rotation periods.   
 
In section \ref{sec:shear} we describe our N-body shearing simulations.
With these simulations in section \ref{sec:mig} we measure
changes in the distributions of guiding centre positions and how they depend
upon time and the strength of  spiral structure in the simulated shearing patch.
In section \ref{sec:oc} we identify a few metal rich young open clusters.
A simple model derived from our simulations is then applied to interpret these
open clusters in terms of constraints on the spiral structure that may have mediated their 
radial migration.
 
\section{Shearing sheet N-body simulations }
\label{sec:shear}

In this section we describe our shearing sheet simulations, dimensions used to characterise them
(section \ref{sec:dim}), how they are set up (section \ref{sec:setup}) and how initial
particle positions and velocities are generated (section \ref{sec:init}).

\begin{figure}
\includegraphics[width=8.5cm]{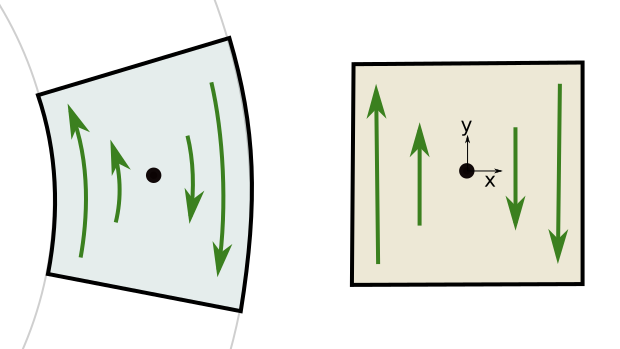}
\caption{An illustration of a patch of a rotating disc (on left) and how
the shearing box (on right) approximates it.
Arrows are shown with respect to motion
in the centre of the disc patch (on left).  In this rotating frame a circular
orbit would remain fixed at the black dot.   An orbit with zero epicyclic amplitude
and located at the centre of the shearing box (on right) would also remain fixed. 
The orientation of our coordinate system is shown on the right.
\label{fig:shearsheet}}
\end{figure}

The shearing sheet \citep{toomre91} approximates a local patch of a rotating disc (see Figure \ref{fig:shearsheet}).
Our N-body simulations of a disc patch use the 
N-body code \texttt{rebound} \citep{rebound} that contains an integrator and shear boundary conditions
 specifically written to simulate a self-gravitating disc patch using the shearing
 sheet approximation \citep{rein12}.  
 From \texttt{rebound} we use the Symplectic Epicycle Integrator (SEI) integrator \citep{rein12}
 and the associated  shear boundary conditions.  
The shear boundary conditions are periodic in two directions, $x,y$ but open in the third direction $z$.
This differs from the simulations by \citet{toomre91} that were restricted to two dimensions.
A square $x,y$ area is simulated that we call the shearing box (see Figure \ref{fig:shearsheet}).
Velocity shear (corresponding to differential rotation) is a function of $x$, so variations in $x$ 
correspond to variations  in galactic radius.  
Variations in $y$ correspond to variations in 
azimuthal angle $\theta$ in the midplane.  
Galactic rotation corresponds to particles
moving  in the $y$ direction.  
The $z$ direction is perpendicular to the disc and shearing box.
Velocity shear itself depend on the parameters $\Omega$ and  $\kappa$
which are independent of position within the shearing box.  Our notation for these
two frequencies follows \citet{B+T,rein12} and  $\Omega$ represents
the angular rotation rate in the disc, and $\kappa$ the epicyclic frequency. 
The parameters $\Omega$ and $\kappa$ are independent of position in the shearing box but
shear is present in the box and the shearing sheet itself approximates a disc that exhibits
radial gradients in both of these frequencies.  The gradient of velocity (the velocity shear corresponding
to differential rotation) depends on both parameters,
as given in equation \ref{eqn:shear}.

Long-range gravitational forces from each particle are computed  using
 the oct-tree  approximation based on  the algorithm by \citet{barnes86}.
Ghost-boxes are used to include forces from particles  that are nearby taking into
account the shearing periodic boundary condition (see section 4.2 by \citealt{rebound}).
Particle collisions are ignored.

The equations of motion in $x,y$ for the shearing sheet are given in the appendix 
(see equations \ref{eqn:motion}).
Orbits in the plane are described by a guiding centre $x_g,y_g$,  epicyclic amplitude $C$
and epicyclic angle $\phi$.  Variations in $x_g$ correspond to variations in guiding radius
or angular momentum in a full disc.  An induced variation in $x_g$ can be called 
radial migration.    Heating or increasing the in-plane components of the velocity dispersion corresponds to
increasing the epicyclic amplitudes (e.g., \citealt{jenkins90}).
Transient spiral structure is expected to cause both heating and migration. However, an individual 
star that migrates a large distance may not be excited to large epicyclic amplitude and the
opposite is also true.

The SEI integrator was written specifically for application in celestial
mechanics  and so has angular rotation rate equal to the epicyclic frequency; $\Omega = \kappa$.  
 We have modified the \texttt{rebound} routines
\texttt{boundaries\_shear.c} and \texttt{integrator\_sei.c}
so that  the epicyclic frequency can take values  $\kappa \ne \Omega$ allowing us
to simulate the shearing sheet corresponding to differential rotation in a galactic disc.
Our code modifications from those described by \citet{rein12} are described in section \ref{sec:app2}.

The equations of motion in $z$ are set by an additional parameter, 
the vertical epicyclic frequency $\Omega_z$ (see section \ref{sec:appz},    
section 3.3 by \citealt{rebound} and
 equation 13 and discussion near this equation by \citealt{rein12}).
However, the actual vertical epicyclic frequency is somewhat faster than $\Omega_z$
due to the self-gravity of the disc.  The value we list in Table \ref{tab:common}  for $\Omega_z$ is 
the parameter set in the code.

\subsection{Dimensions}
\label{sec:dim}

The natural unit of time for the shearing sheet is  $\Omega^{-1}$, 
or the associated orbital period $P\equiv 2\pi/\Omega$.  
With a self-gravitating disc of mean mass surface density $\Sigma$,
a natural unit of distance
is the Toomre wavelength, often called the critical wavelength,
\begin{align}
\lambda_{crit} & \equiv \frac{4 \pi^2  G \Sigma}{\kappa^2} \nonumber \\
&=1 {\rm kpc} \left(\frac{\Sigma}{10 M_\odot{\rm pc}^{-2} }\right)
\left(\frac{ 2}{\kappa^2/\Omega^2} \right)
\left(\frac{\Omega}{30\ {\rm km\ s}^{-1}{\rm kpc}^{-1} }\right)^{-2}, \label{eqn:lambda_crit}
\end{align}
where $G$ is the gravitational constant.   This wavelength
is independent of the particle velocity dispersion.
Where there are variations in the surface density $\Sigma$ we use the mean of the surface
mass density distribution, $\mu_\Sigma$,
to compute $\lambda_{crit}$.
For comparison,
the estimated total stellar (including  brown dwarfs, white dwarfs and other remnants) 
 surface density in the solar neighbourhood
is $\Sigma_* \approx 33 M_\odot {\rm pc}^{-2}$ 
and in gas $\Sigma_g \approx 14 M_\odot {\rm pc}^{-2}$ 
with $ \approx 7 M_\odot {\rm pc}^{-2}$ in cold molecular and atomic hydrogen
near the midplane \citep{mckee15}.
The Galactic disc baryonic components  total just under $50 M_\odot {\rm pc}^{-2}$, 
similar to previous estimates for the local disc surface density \citep{flynn06}.

Swing amplification is strongest at about the Toomre wavelength 
so a self-gravitating
disc most quickly grows spiral structure of this wavelength \citep{toomre81,ath84,toomre91,fuchs01}.
We set the box size of the shearing sheet simulation to exceed the Toomre wavelength (as did
\citealt{toomre91}) so that
the simulations can resolve this wavelength.  
 
The Toomre $Q$-parameter \citep{safronov60,toomre64} for a stellar disc 
depends on the stellar velocity dispersion with $\sigma_{vR}$
the standard deviation of its radial component,   
\begin{align}
Q &\equiv \frac{\sigma_{vR} \kappa}{3.36 G \Sigma} \nonumber \\
&=1.1 \left( \frac{\sigma_{vR}}{20\ {\rm km~s}^{-1}} \right) 
\left( \frac{\kappa/\Omega}{\sqrt{2}}\right) \times \nonumber \\
& \qquad \qquad
\left(\frac{\Omega}{30\ {\rm km\ s}^{-1}{\rm kpc}^{-1}} \right)
\left(\frac{\Sigma}{50 M_\odot{\rm pc}^{-2} }\right)^{-1}. \label{eqn:QToom}
\end{align}
The amplitude or strength of spiral structure is primarily set by the Toomre $Q$-parameter \citep{fujii11}.
For the shearing sheet simulations we use only the massive particles
to compute the standard deviation 
of the $x$ velocity component
$\sigma_{vx}$, and replace $\sigma_{vR}$ with $\sigma_{vx}$ to compute the Toomre $Q$-parameter.

\subsection{Simulation set up and drag force}
\label{sec:setup}

We use massive particles to generate self-gravitating spiral structure.   
A thousand massless  particles embedded within the simulation
are used as tracers to track variations in guiding centre position $x_g$.
Spiral structure induced drifts in tracer particle $x_g$ values are interpreted as
 radial migration.    
Tracer particles are point masses and so can represent stars or compact
star clusters.   
In section \ref{sec:oc} we use the results of our shearing sheet
simulations to  discuss migration  of open clusters, 
assuming that they 
do not strongly perturb the background Galactic spiral structure
and neglecting processes of cluster evaporation and dissolution.
We assume that the size of the shearing box significantly exceeds the
size of a star cluster.

A disc initially set with Toomre $Q \gg 2$ will not show spiral structure, whereas one initially set
with $Q\la 1$ will be unstable, allowing spiral structure to grow.
As a simulation runs, the spiral structure itself heats the disc and increases the Toomre $Q$-parameter.
Growth and decay of transient spiral structure
induces variations in the epicyclic amplitude of stars
\citep{carlberg85,jenkins90} and this can happen over a range of radius as
stars need not be near a Lindblad resonance.
The heating rate is faster at low Toomre $Q$-parameter values \citep{fujii11}.  To allow us to run simulations
that show high amplitude spiral structure but only slowly vary in Toomre $Q$-parameter,
we added a small fictitious drag force to the massive particles as an additional force to \texttt{rebound} so
as to cool the disc.  Our drag force is identical to that used by \citet{toomre91} (see the top of their page 350),
 is a straightforward velocity dependent acceleration in the $x$ direction 
\begin{equation}
a_x = - \alpha \Omega v_x, 
\end{equation}
and  is described by a single unitless parameter $\alpha$ which 
sets the timescale for damping the epicyclic amplitude. 
We applied the drag force to massive particles only
and only in the $x$ direction so we did not need to take into account 
the velocity shear to compute a damping force and so that vertical motions are unaffected.
At larger values of Toomre $Q$-parameter and with weaker spiral structure, the heating rate
is reduced and so damping was not necessary to maintain a slowly varying Toomre $Q$-parameter. 
Our damping serves as a source of dissipation, replacing hydrodynamic dissipation that
 would be modeled in more realistic simulations
that include a gaseous as well as a stellar disc component.
\citet{toomre91} verified that a steady state in the velocity distribution could be
reached after many orbits of integration.\footnote{At large values of $\alpha$ and 
with small softening lengths, clumps can form in the disk (Agris Kalnajs, private communication).}  
Our goal is to look at the extent of migration over only a few orbital periods so it is not 
necessary
to maintain exact stability in the character of spiral structure over many orbital periods. 
However the heating and migration rate are affected by slow variations in the spiral
structure and these are present in the simulations.   
We will discuss this sensitivity later when we study variations in the distributions.

We first choose a number of massive particles
to simulate and a mean disc surface density, $\Sigma$.  
The shearing box size is chosen to exceed the Toomre wavelength.
The massive particle masses are identical  and set using the box size and mean disc surface 
density.  

Common parameters for our simulations are listed
in Table \ref{tab:common} and those for individual simulations in Table \ref{tab:simlist}.
The rotation curve near the solar neighborhood is nearly flat (see section 6.4 and Figure 16
by \citealt{bland16}) corresponding to $\kappa/\Omega \approx \sqrt{2}$.
The value chosen for this ratio (and listed in Table \ref{tab:common}) is approximately
consistent with the differential rotation of a flat rotation curve.

While it is natural to work with time in orbital periods
$P = 2 \pi /\Omega$ and length in 
units of the Toomre wavelength, it is helpful for interpretation to relate these
to actual physical units.    
The angular rotation rate
near the Sun is about $\Omega \sim 30$ km s$^{-1}$ kpc$^{-1}$ corresponding to an orbital period 
of about 200 Myr (using values by \citealt{bland16}).
Through-out this manuscript we give lengths in units of Toomre wavelength
and in pc for a Toomre wavelength of $\lambda_{crit,0} = 1007$ pc computed for a 
mass surface density of $\Sigma_0 = 10 M_\odot {\rm pc}^{-2}$.
Equation \ref{eqn:lambda_crit} can be used to estimate distances for another 
value of mean surface density by multiplying the Toomre wavelength
by the desired $\Sigma$ divided by the value $\Sigma_0 = 10 M_\odot {\rm pc}^{-2}$.
When working in pc,  Myr and solar masses,  velocities are in pc/Myr $\sim $ km/s and
the gravitational constant
$G = 0.0045 M_\odot^{-1} {\rm pc}^{3} {\rm Myr}^{-2}$.

\subsection{Initial conditions for particles}
\label{sec:init}

For the massive particles,  initial guiding centre coordinates
$x_g, y_g$  are chosen randomly using uniform probability distributions covering the area of the shearing box.
The in-plane and vertical epicyclic angles are randomly chosen from
 uniform probability distributions in $[0,2\pi]$.  The in-plane and vertical epicyclic amplitudes are
randomly chosen from uniform probability distributions ranging from zero to maximum values.
Initial particle positions and velocities are computed from the epicyclic amplitudes and angles
and guiding centre coordinates using equations \ref{eqn:motion} and \ref{eqn:zvz}.
The resulting massive particle distribution is uniformly distributed in $x,y$ in the shearing box.
So there is no gradient in the mean mass surface density $\Sigma$ in the shearing sheet.
The maximum value for the vertical epicyclic amplitude sets the disc thickness,
whereas the maximum in-plane epicyclic amplitude sets the initial Toomre $Q$-parameter.
The number of massive particles and smoothing or gravitational softening 
length were chosen to be large enough that the simulations
are not highly sensitive to either value.    We will illustrate how variations in these quantities
affect our results in section \ref{sec:checks}.  The 2 dimensional shearing sheet
simulations by \citet{toomre91}
used a larger smoothing length than ours ($\sim 0.2 \lambda_{crit}$) 
perhaps in part to mimic the behavior of disk thickness.

Massless tracer particles
are used to measure variations in guiding radius $x_g$ corresponding to migration.
After two rotation periods, the growth rate of spiral structure in the massive
particles is reduced. 
The tracer particles are only added to the simulation after two orbital
periods, after which time the amplitude of spiral structure varies less quickly.  
Tracer particles are added after spiral structure is grown so as to mimic the birth of
stars and clusters into a galaxy in which spiral structure is present.
After the 1000 tracer particles are added, 
the simulation is  integrated for 5 additional orbital periods.
Our figures show time in units of orbital periods from the time when the tracer  particles
are added to the simulation.

\begin{figure*}
\includegraphics[height=4.0cm, trim={22mm 0 15mm 0},clip]{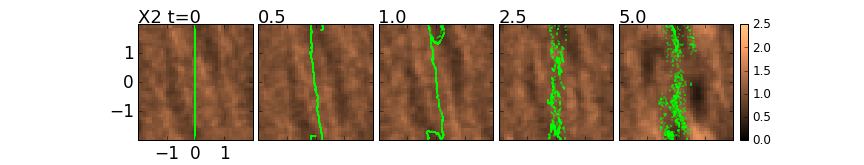}
\caption{Surface mass density of massive particles shown as an image 
with the positions of 1000 massless tracer 
particles, shown as green dots,  at 5 different times in the X2 simulation.
The times for each snapshot are 0, 0.5, 1.0, 2.5 and 5.0 orbits (from left to right)
after the tracer particles are injected into the simulation.
The colour range displayed  is 0 to 2.5 times the mean mass surface density.
All panels have the same colour display range.
The $x$ and $y$ axes are in units of the Toomre wavelength and the entire shearing box is shown.
The simulation parameters are listed in Tables \ref{tab:common} and \ref{tab:simlist}.
\label{fig:line}}
\end{figure*}

Tracer particles are begun in the plane $z=0$ and at $x =0$ 
but with $y$ values chosen from a uniform distribution covering the width of the shearing box.
The initial distribution can be seen in the leftmost  panel in Figure \ref{fig:line}.
The velocity is set to zero so the particle initially has guiding radius $x_g = 0$,
epicyclic amplitude $C=0$ and zero vertical epicyclic amplitude.    
In the absence of spiral perturbations the tracer particles
would remain fixed  (see equations \ref{eqn:motion} and\ref{eqn:shear}).
As the tracer particles are  massless,  their initial linear distribution does not disturb
the development and evolution of spiral structure.
Because tracer particles are begun with $x_g=0$, the absolute value of the 
guiding position $|x_g(t)|$ is an estimate for the distance migrated.   
As our tracer particles are point masses, the sizes of 
the clusters that they represent  are neglected.
The migration distances we consider
 are similar to or greater than hundreds of pc and so we neglect the much smaller initial cluster
size (about 1 pc).

\begin{figure*}
\includegraphics[height=4.0cm, trim={22mm 0 15mm 0},clip]{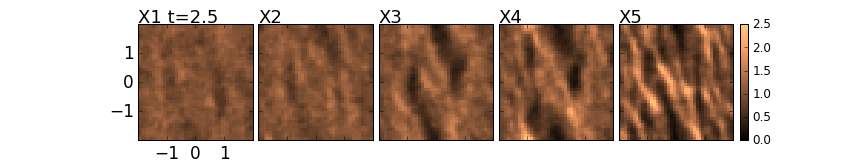}
\caption{For five different simulations we show the mass surface density 
at $t=2.5$ orbits after tracer particles are
injected into the simulation. 
The colour range displayed  is 0 to 2.5 times the mean density.
All panels have the same colour display range.
The $x$ and $y$ axes are in units of the Toomre wavelength and the entire shearing box is shown.
The Toomre $Q$-parameters for these simulations are measured at 2.5 orbital periods
after tracer particles are added to the simulation.
From left to right we show high Toomre $Q$ to low Toomre $Q$-parameter simulations in the X series (X1--X5).
The simulation parameters are listed in Tables \ref{tab:common} and \ref{tab:simlist}.
\label{fig:dens}}
\end{figure*}

\begin{figure*}
\includegraphics[height=4.0cm, trim={22mm 0 15mm 0},clip]{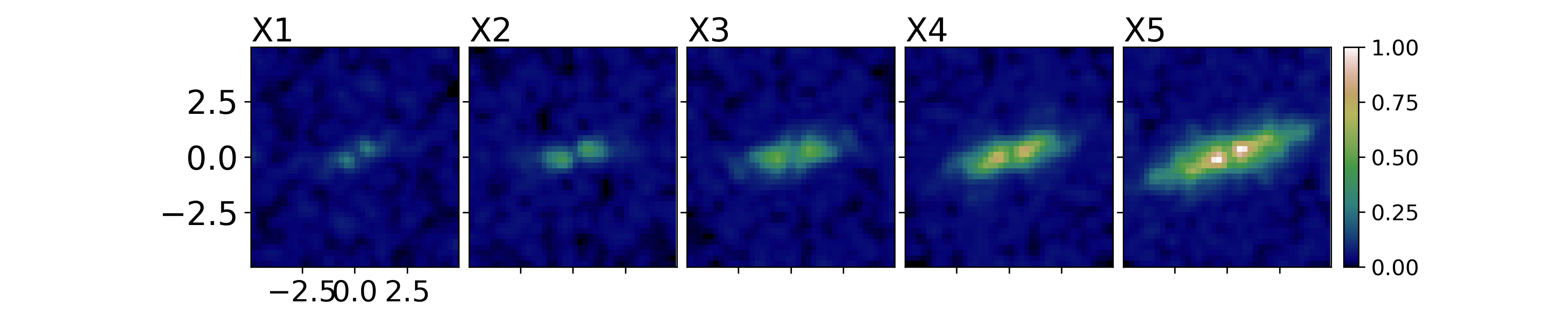}
\caption{2D Fast Fourier transforms of each of the mass surface densities  
shown in Figure \ref{fig:dens} (at 2.5 orbital periods) were computed from massive particles after subtracting the mean mass surface density.  
The images show  Fourier amplitudes and
all panels have the same colour display range.  
The colour bar scale is set so that a sine wave with amplitude equal to the mean density
gives power of amplitude 1.
The centre of the image contains low frequency power.  
The maximum spatial frequencies (on the boundaries of each image) are  
$4.91 \lambda_{crit}^{-1}$ (or 0.0049 pc$^{-1}$) corresponding to 
wavelengths of  $0.2 \lambda_{crit}^{-1}$ (208 pc for $\lambda_{crit,0}  = 1007 $ pc).
The angle of the power distribution seen in these 2D spectrograms depends on the angle of the spiral features.
The lower Toomre $Q$-parameter simulations (on the right) have more power than the higher Toomre $Q$-parameter simulations.
The lowest Toomre $Q$-parameter simulations have  broad
spatial frequency distributions containing power on short and long spatial wavelengths.
\label{fig:fft}}
\end{figure*}

\begin{table}
\vbox to90mm{\vfil
\caption{\large  Common simulation parameters \label{tab:common}}
\begin{tabular}{@{}lllllll}
\hline
Number of massive particles   &   50000  \\
Number of massless tracer particles  & 1000     \\
Time when tracers are added to simulation & 2 orbital periods \\
Integration time after tracers are added & 5 orbital periods \\
Time step  & 0.002 orbital periods \\
$\kappa/\Omega$ & 1.4 \\
$\Omega_z/\Omega$ & 1.8 \\
Smoothing length   &   0.0496 $\lambda_{crit}$ = 50 pc\\
Shearing box length &  $3.97 \lambda_{crit}$   = 4.0 kpc\\
Massive particle mass & $3200 M_\odot $ \\
$\sigma_z$  & $0.15 \lambda_{crit}$  = 150 pc\\
\hline   
\end{tabular}
{\\   Notes.  In the above Table, dimensions in pc and $M_\odot$ are given for a disc
with $\Sigma_0 = 10\ M_\odot {\rm pc}^{-2}$ and Toomre wavelength $\lambda_{crit,0} = 1007$ pc.
Here $\sigma_z$ is the standard deviation of $z$ for the massive particles. 
The circular velocity around one particle at a smoothing length is 0.5 km/s, computed with
$G = 0.0045 M_\odot^{-1} {\rm pc}^{3} {\rm Myr}^{-2} $.
}}
\end{table}

\begin{table}
\vbox to90mm{\vfil
\caption{\large  List of simulations \label{tab:simlist}}
\begin{tabular}{@{}lllllll}
\hline
Simulation & $Q$  & $\alpha$ & $\sigma_\Sigma/\mu_\Sigma$ &differences  \\
\hline
X1 & 2.4 &0.005  & 0.20 & \\
X2 & 2.0 &0.005  & 0.22 & \\
X3 & 1.6 &0.005 &  0.26 &\\
X4 & 1.4 & 0.02 &  0.34 &\\
X5 & 1.2 &0.05 & 0.46 & \\
\hline
X3S  &1.6 &0.005& 0.26 & smaller smoothing length \\
X3N  &1.6 &0.005 & 0.32 & fewer particles \\
X3ha &1.6 & 0.005 & 0.24 & thicker disc \\
X3hb &1.7 &0.005 & 0.30 & thinner disc \\
\hline   
\end{tabular}
{\\  Notes.  
The X3S, X3N, X3ha, X3hb simulations are similar to the X3 simulation
except as described in the rightmost column.
X3S has half X3's smoothing length, X3N has half X3's number of particles,
X3ha has twice X3's disc thickness and X3hb has half X3's thickness.
The Toomre $Q$-parameter is measured at 2.5 orbital periods
after tracer particles are added to the simulation.
The ratio of the standard deviation to mean of the surface density distribution
$\sigma_\Sigma/\mu_\Sigma$ is computed at the same time and
 is a measure of surface density contrast.
 The drag force for massive particles is set by $\alpha$.
}}
\end{table}

\FloatBarrier

\section{Migration on the shearing sheet}
\label{sec:mig}

After listing our simulations,
we discuss in section \ref{sec:snap} the morphology of spiral structure.
In section \ref{sec:dist} we show guiding centre distributions
as a function of time,  illustrating spiral structure induced radial migration.
As tracer particles migrate away from their birth positions, their guiding centre distributions widen.
In section  \ref{sec:sig} the standard deviations of these distributions are shown.
In section \ref{sec:checks} we discuss numerical checks on the code.
In sections \ref{sec:1orb} and \ref{sec:time} we fit functions to the standard deviations 
of the guiding centre distributions.
Maximal migration distances as a function  of time are discussed in section \ref{sec:max}.


We ran a series of simulations with shearing box size approximately 4 times 
 the Toomre wavelength.  Five simulations X1--X5 are run with identical parameters
except with differing initial in-plane velocity dispersions for the massive particles
and different levels of damping, $\alpha$ (see Tables \ref{tab:common}, \ref{tab:simlist}).
The values of damping parameter 
$\alpha$ imply that damping for massive particles is slow, even for the low $Q$ simulations.
Particle positions and velocities are output every 0.5 orbital periods.

The Toomre $Q$-parameters  are measured 2.5 orbits after the tracer particles
are added to the simulation and these too are listed in Table \ref{tab:simlist}.
Four additional simulations were run.  The X3S simulation is identical to the X3
simulation except the smoothing length is half the size of that listed in Table \ref{tab:common}.
The X3N simulation is identical to the X3 simulation except it has only 25000 massive
particles instead of 50000.   The X3ha and X3hb simulations are identical to the X3
simulation except X3ha has a vertical standard deviation, $\sigma_z$, twice that of X3
and X3hb has a vertical standard deviation half that of X3.
The vertical standard deviations are computed from the $z$ distributions of massive particles.

\subsection{Simulation snapshots}
\label{sec:snap}

The surface mass density of massive particles along with the positions of the 1000 massless 
tracer particles are shown in Figure \ref{fig:line}
 at 5 different times in the X2 simulation.
The times for each snapshot are 0, 0.5, 1.0, 2.5 and 5.0 orbits 
after the massless tracer particles are injected into the simulation.
The leftmost panel shows that spiral structure has grown prior to the insertion of
our massless tracer particles.   The tracer particles are inserted at $x=0$ where
there is no drift in guiding centre position.  Without spiral structure each point in the vertical green line in 
the leftmost panel of Figure \ref{fig:line}
would remain fixed.  The velocity shear is such that the right hand side of the box moves
downward and the left hand side of the box moves upward.

One half an orbit later (second panel from left in Figure \ref{fig:line})
the green line has become wavy as the tracer particles have been perturbed by nearby spiral
structure.  Perturbations excite epicyclic motions as well as move guiding
centres so the width of the $x$ distribution is only approximately
equivalent to the width of the distribution of $x$ component of the guiding centre distribution.
By the end of the simulation (rightmost panel) the tracer particles
have become dispersed. 
We do track boundary crossings for the tracer particles in case migration is extensive.
However, with our shearing box length exceeding
the Toomre wavelength and within 5 orbital periods, we saw no shear box boundary crossings
in the $x$ direction.  Tracer particles did not cross from the right hand side to the left or vice versa.
Tracer particles do cross from the top to the bottom boundary (and vice versa) due to the velocity
shear.

A comparison between the leftmost three panels in Figure \ref{fig:line} show that the spiral structure
has some coherence over an orbit. However spiral arms vary (as a function of time) in position
and amplitude or strength.  
A difference between a shearing sheet simulation and an
N-body simulation of a full disc is that in the shearing sheet there cannot be coupling of patterns
from one radius to another.  All spiral features are nearly corotating
with the background velocity shear (this is also discussed by \citealt{toomre91}). 
We have verified this by
plotting density slices from the shearing box versus time.  Using particles in the centre of the image (at $x=0$)
we construct a density histogram giving densities as a function of $y$ and $t$.
There is little structure in this density histogram image 
as expected for patterns moving at corotation with the velocity shear in the box.
Likewise, 
at $x<0$ or $x>0$ in a $y$ vs $t$ density histogram image
we  do see streaks due to the velocity shear.    Bumps in the density field
from spiral arms move approximately with the background shear velocity field, as would
be expected from corotating patterns.

A comparison between the leftmost and rightmost panel in Figure \ref{fig:line} 
shows that the spiral structure
is not uniform across the five orbits.  The Toomre $Q$-parameter does change 
across the simulation (ranging from 1.3 to 1.6), and
the spiral structure has higher density peaks and larger wavelength
at later times.   
Our procedure for damping particles has not completely stabilized the disc.  We attribute
the slow evolution to the slower growth of wavelengths that differ 
from the peak wavelength favored by swing amplification.  We keep in mind that
slow variations in spiral structure in the simulations  make it more
 difficult to predict properties
of the distributions of the $x$-component of the guiding centre for the tracer particles as a function of time.

In Figure \ref{fig:dens} we show the surface density distribution for X1--X5 simulations
in order of decreasing Toomre $Q$-parameter (from left to right) but at $t= 2.5$ periods  after
the tracer particles are inserted into the simulation.
These snapshots illustrate that spiral structure is higher amplitude 
for the low Toomre $Q$-parameter simulations.  
The variation in density contrast between simulations is larger than the slow drifts during each 
individual simulation.
Even though there are slow drifts in Toomre $Q$-parameter
and spiral morphology across each
 simulation, there should be large differences in the migration rates of
 the tracer particles as the simulations span a large range in spiral amplitude.

We compute  two-dimensional Fourier transforms of the images shown in Figure \ref{fig:dens}
and show the Fourier amplitudes in 
Figure \ref{fig:fft}.   Our figure \ref{fig:fft} resembles the similarly computed Figures 3 and 4 
by \citet{toomre91}  for their shearing sheet simulations.  These were interpreted
as showing particle-particle spatial correlations due to spiral {\it wakes} \citep{julian66}
and are caused by amplification of small over-densities by self-gravity \citep{julian66,toomre81}.  
There is more power  in the lower Toomre $Q$-parameter simulations than
the high Toomre $Q$-parameter simulations.  The lowest Toomre $Q$-parameter
 simulation (rightmost panel) has a much broader
spatial frequency  distribution containing power on short and long spatial wavelengths.
Hence the spiral structure is not restricted to a single wavelength and a single amplitude
associated with it.  If we used a low order Fourier decomposition to model the spiral structure
we would likely  underestimate heating (in epicyclic amplitude) and migration rates.

\subsection{Distributions of guiding centres}
\label{sec:dist}

To characterize migration we measure the distribution of the $x$ component of
the guiding centre, $x_g$, for massless tracer particles as a function
of time.  The $x$ component of the guiding centre is computed using equations \ref{eqn:xg} 
from tracer particle positions and velocities.  
As tracer particles initially all have $x_g=0$, the distributions at later times are sensitive to
the extent of migration.    
Spiral structures could cause the guiding centre $x_g$ 
of a particle to oscillate about a mean value (e.g., see Figure 4 by \citealt{comparetta12}).

We consider
migration to be a drift in the $x_g$ mean value,   ignoring
short period oscillations about this mean, however both
short timescale oscillations and longer timescale drifts would  affect
the guiding centre distributions.  We assume
that the distributions are dominated by the slow drifts and so illustrate migration, though
the  short timescale oscillations  could 
affect the distributions at early times.
Our initial tracer particle distribution  is a delta function at $x_g=0$.  A wider initial distribution can be
considered a sum of narrow spikes each with a different initial $x_g$.
The distribution in $x_g$ at a later time for a wider initial distribution can be estimated by convolving
the  distribution we find at the same later time (derived from our initially narrow distribution) with the function
describing the initial wider $x_g$ distribution.

\begin{figure}
\includegraphics[width=3.5in]{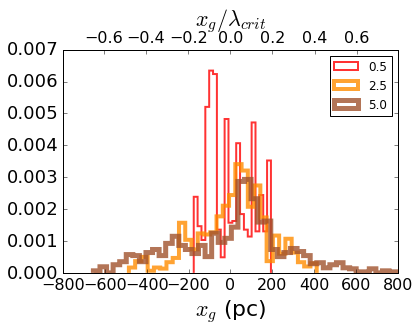}
\caption{Distribution of the $x$ component of the tracer particle guiding centres, $x_g$, 
at three times in the X3 simulation.   
All tracer particles had initial $x_g=0$, as shown in Figure \ref{fig:dens}.
The thin red line is at $t=0.5$ orbital periods, the mid-weight orange line at 2.5 orbital
periods and the thick brown line at 5 orbital periods.
The $x$ axis is shown in units of pc for a mean surface density
$\Sigma_0 =10 M_\odot {\rm pc}^{-2}$  (bottom axis) and in units of Toomre wavelength
$\lambda_{crit}$ (top axis).
The distributions are normalized so that they integrate to 1.
The width of the distributions increases in time.  At later times  the distributions are smooth enough 
that a diffusive approximation might be valid, despite the
short timescale. 
\label{fig:dist_xg}}
\end{figure}

Guiding centre distributions are shown for the X3 simulation
at 0.5, 2.5 and 5 orbital periods after tracer particle insertion in Figure \ref{fig:dist_xg}.
The distributions are normalized so they integrate to 1.
Individual spikes at early times are likely caused by individual spiral features, with 
the distributions becoming smoother at later times. These were also noted by
\citet{toomre91} who described them as guiding centre `bunchings.'   Below we  measure
the standard deviations of these guiding centre distributions but will refer to this figure later to
discuss the tails of the distribution.  The tails are relevant for estimating
how far a particle can get from its birth guiding centre radius.

Figure \ref{fig:dist_xg} shows that 
 guiding centre distributions broaden in
only  five  rotation periods.   The original spiral heating \citep{carlberg85,jenkins90} and migration models
\citep{sellwood02} were mediated by growth and disappearance of individual spiral patterns.
If the growth and disappearance of a spiral pattern requires a few orbital
periods then within  five rotation periods there is only time 
for one or two patterns to appear and disappear.  
The guiding centre distributions 
are smooth enough at later times that they could be  consistent with a diffusive
model, valid in the limit where perturbations to the $x$-component of the guiding centres occur randomly and
many times, not just once or twice.  The diffusive behavior can be reconciled with the short timescale
if individual spiral features are uncorrelated or if patterns interfere with one another 
(as proposed by \citealt{comparetta12}).   A re-examination of the simulations
by \citet{toomre91} suggest that each swing amplified
structure, seen by growth and variation in Fourier amplitude, also moves the guiding centers of groups of particles\footnote{Agris Kalnajs, private communication}.   
Stochastic variation in guiding centers may be a local process associated
with swing amplification of weak density variations that are present in our simulations
because of numerical noise associated with the finite particle number, but also present in the Galaxy
from molecular clouds and star clusters.  

\subsection{Broadening of the guiding centre distributions}
\label{sec:sig}

\begin{figure}
\includegraphics[width=3.5in]{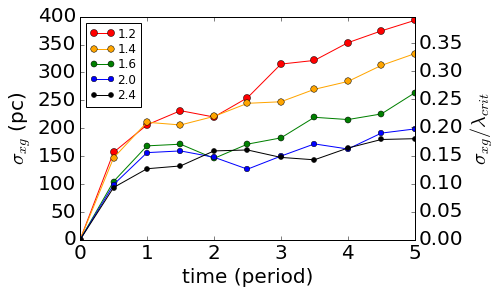}
\caption{
We show the standard deviation $\sigma_{xg}$  of the $x$ component of the tracer particle 
guiding radii   as a function of time for X1--X5  simulations
also shown in Figure \ref{fig:dens}.
Higher Toomre $Q$-parameter simulations have less and slower radial migration.
The simulations are labelled by their Toomre $Q$-parameter value mid-simulation (see Table \ref{tab:simlist}).
From top to bottom the simulations are  X5 (red points), X4 (orange), 
X3 (green), X2 (blue) and X1 (black).
The $y$ axis is shown in units of pc for a mean surface density
$\Sigma_0 =10 M_\odot {\rm pc}^{-2}$  (left axis) and in units of Toomre wavelength
$\lambda_{crit}$ (right axis).
\label{fig:sig_xg}}
\end{figure}

We measure the standard deviation $\sigma_{xg}$ of the $x$ component of the guiding centre, $x_g$,
for  the tracer particles as a function of time
and these are plotted for the X1--X5 simulations in Figure \ref{fig:sig_xg}.
The standard deviations characterize the width of the distributions
shown in Figure \ref{fig:dist_xg}. Even for the strongest spiral structure (the X15simulation), 
within 5 orbital periods
the standard deviation in $x_g$ remains less than half a Toomre wavelength.

Figure \ref{fig:sig_xg} shows that the distributions of the $x$ components of the guiding centres 
 rapidly spread within the first
orbital period.  The rapid growth is likely because our tracer particles were begun
in circular orbits and inserted abruptly into a simulation with spiral structure.
We  experimented with starting our tracer particles with velocities near those
of massive particles (moving with the spiral structure) or starting them at the beginning
of the simulation with the massive particles but saw similar 
  standard deviations early in the simulation.
The X3 simulation reaches a maximum of $\sigma_{xg}/\lambda_{crit} \approx 0.26$
and this can be compared to the maximum reached by a single tracer particle (out of 1000) of $\sim 0.7$
as shown in Figure \ref{fig:dist_xg}.  

\begin{figure}
\includegraphics[width=3.5in]{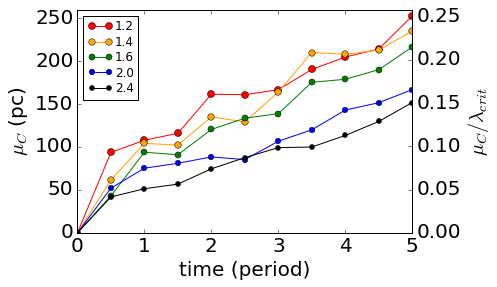}
\caption{Mean epicyclic amplitude for tracer particles as a function of time for the X1--X5 simulations.
The simulations are labelled by their Toomre $Q$-parameter value mid-simulation with points as in 
Figure \ref{fig:sig_xg}.  
The $y$ axis is shown in units of pc for a mean surface density
$\Sigma_0 =10 M_\odot {\rm pc}^{-2}$  (left axis) and in units of Toomre wavelength
$\lambda_{crit}$ (right axis).
\label{fig:sig_A}}
\end{figure}

Figure \ref{fig:sig_xg} shows that the width of the guiding centre distributions 
 depends on the Toomre $Q$-parameter even at early times in the simulation. 
 The rate that the standard deviation of the distribution $\sigma_{xg}$ increases
is large at the beginning then decreases  past one orbital period.  
The rates that $\sigma_{xg}$ increases 
 past $t=1$ period are shallower for the higher Toomre $Q$-parameter simulations.
We discuss possible explanations for this.
As particles increase in epicyclic amplitude, they may be less likely to
be migrated by corotating spiral features \citep{daniel15}.
 To explore this possibility,
mean epicyclic amplitudes for the same simulations as a function of time are
shown in Figure \ref{fig:sig_A}.  The epicyclic amplitudes are not large enough that multiple spiral features
are crossed by  single particles in their orbits. 
Figure \ref{fig:sig_A} shows that tracer particles in the lower Toomre $Q$-parameter simulations have higher
epicyclic amplitudes.  The lower Toomre $Q$-parameter simulations have stronger
spiral structure and so would exhibit increased heating \citep{carlberg85,jenkins90}.   If the extent of migration 
decreases as the epicyclic amplitudes increases 
(as the probability for capture into resonance decreases; see section 2.3 by \citealt{daniel15}),
the slope  in $\sigma_{xg}(t)$ for the low Toomre $Q$-parameter simulations would be shallower
rather than higher as seen in Figure \ref{fig:sig_xg}.   The trend is opposite
to that expected if the slope variation is due to a decrease in migration rate caused
by an increase in epicyclic amplitude.\footnote{In contrast re-analysis of the
the shearing sheet simulations by \citet{toomre91} shows that the guiding center distribution
standard deviations are insensitive to initial particle epicyclic amplitude; Agris Kalnajs, private
communication.} 
 
A second possible explanation for the steeper slopes at lower Toomre $Q$-parameter past $t=1$ period
(seen in Figure \ref{fig:sig_xg}) is  that
the higher Toomre $Q$-parameter simulations have 
slower variations in spiral morphology (amplitudes and  pattern speeds)
 than the lower Toomre $Q$-parameter simulations.
If the spiral amplitudes increase more rapidly in the low Toomre $Q$-parameter simulations then
the migration rate also would increase throughout the simulation.
The most likely explanation for the differences in slope at later times in Figure \ref{fig:sig_xg}
are differences in time dependent spiral structure morphology. 

\subsection{Sensitivity to disc thickness}
\label{sec:checks}

\begin{figure}
\includegraphics[width=3.5in]{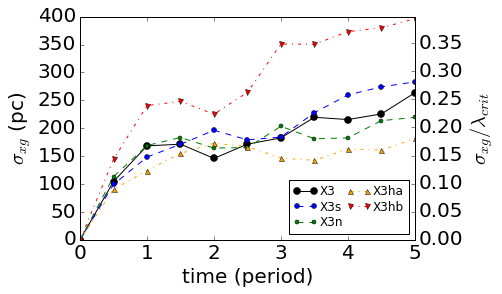}
\caption{
We show the standard deviation $\sigma_{xg}$  of the $x$ component of the tracer particle
guiding radii  for the simulations X3, X3S, X3N, X3ha, and X3hb.
The standard deviations are not strongly sensitive to the smoothing length (comparing X3S to X3)
or number of
particles (comparing X3N to X3), but are sensitive to the disc thickness (comparing X3ha, X3hb to X3). 
The $y$ axis is shown in units of pc for a mean surface density
$\Sigma_0 =10 M_\odot {\rm pc}^{-2}$  (left axis) and in units of Toomre wavelength
$\lambda_{crit}$ (right axis).
\label{fig:sig_xg_comp}}
\end{figure}

Before we explore models for the time dependence of the
guiding centre distributions we check the sensitivity of the
simulations to vertical thickness, particle number and smoothing length. 
Figure \ref{fig:sig_xg_comp} shows standard deviations (of $x_g$) for 4 simulations that
are similar to the X3 simulation (see Table \ref{tab:simlist}).
Compared to the X3 simulation,
the X3S simulation has half the smoothing length, the X3N simulation
has half the number of massive particles, and the X3ha and X3hb simulations have twice and half
 as thick discs.
Figure \ref{fig:sig_xg_comp} shows that our simulations are not 
strongly sensitive to the smoothing length (comparing X3S to X3)
or number of particles (comparing X3N to X3). 
 However the migration rates
are sensitive to the disc thickness, with the thinner disc (X3hb) having
more extensive migration.  
The sensitivity of the standard
deviations to disc thickness will be discussed further at the end of section \ref{sec:1orb}.

Up to this point we have discussed
simulations with tracers initially placed in the midplane.
We ran a simulation similar to the X3 simulation but with tracer particles begun
with the same vertical dispersion ($\sigma_z$) as the massive particles.  
The standard deviations of $x_g$ displayed
no  significant differences compared to the X3 simulation.   

\subsection{Guiding centre standard deviations at 1 orbital period}
\label{sec:1orb}

Our simulation snapshots (Figures \ref{fig:line}, \ref{fig:dens}) and Fourier amplitudes
(Figure \ref{fig:fft}) show that the simulations are poorly described by a single spiral wavelength.
Measurements of the peak mass surface density in the sheet as a function of time
wildly fluctuated, possibly because of interference
between spiral features. 
For a more robust measurement of spiral strength 
 we use the standard deviation of the mass surface density distribution divided
by the mean surface density $\sigma_\Sigma/\mu_\Sigma$ and refer to this quantity
as the surface density contrast.  It 
 is 0 for a uniform surface density disc and increases with the strength of spiral structures.
The surface density contrasts are measured for each simulation at $t=2.5 $ orbits
 and listed in Table \ref{tab:simlist}.

We found that the surface density contrast
increases with decreasing Toomre $Q$-parameter  in the X1--X5 simulations, as expected.
Plotting this against the standard deviation $\sigma_{xg}$ revealed a trend, 
similar to that found by \citet{fujii11} for the dependence of heating rate
on Toomre $Q$-parameter and spiral amplitude.  However the trend was not matched by the thick and thin
disc simulations (X3ha, X3hb) until we also included a correction for disc thickness.
This approach was also explored
by \citet{fujii11} for disc heating.

\begin{figure}
\includegraphics[width=3.0in]{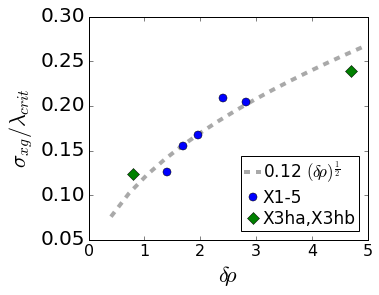}
\caption{The standard deviation $\sigma_{xg}/\lambda_{crit}$ in guiding radius  
for tracer particles versus midplane density contrast $\delta \rho$ (equation \ref{eqn:deltarho}) 
for  simulations X1--X5, X3hb, and X3hb, computed at a time of 1 orbit.
The green diamonds  show the thick and thin disc (X3ha, X3hb) simulations, whereas
the blue circles  are simulations X1--X5.  The initial increase in guiding radius scales with density contrast.
The dashed grey line shows the function in equation \ref{eqn:yy}.
\label{fig:trend_rho}}
\end{figure}

The trend is shown in Figure \ref{fig:trend_rho} where we plot standard deviation
$\sigma_{xg}/\lambda_{crit}$ against a unitless form for the
 density contrast  in the midplane
\begin{equation}
\delta\! \rho \equiv  \frac{\sigma_\Sigma}{\mu_\Sigma}\frac{\lambda_{crit}}{\sigma_z}. \label{eqn:deltarho}
\end{equation}
In Figure \ref{fig:trend_rho}
 both quantities  ($\delta\! \rho$ and $\sigma_{xg}/\lambda_{crit}$) 
 are computed 1 orbit after the tracer particles are inserted into the simulation.
As $\sigma_\Sigma/\mu_\Sigma$ is a measure of the surface density contrast
and $\sigma_z/\lambda_{crit}$ characterizes the thickness of the disc, their ratio
characterizes spiral feature density contrast.
Figure \ref{fig:trend_rho}  shows that the width of the distribution
of the $x$-component of the guiding centres  
is related to the midplane density contrast due to  spiral structure.
The dot dashed line in Figure \ref{fig:trend_rho} shows the curve 
\begin{equation} 
f(\delta\! \rho) = 0.12 \sqrt{\delta\! \rho} \label{eqn:yy}
\end{equation}
which captures the trend seen with the points.
The coefficient 0.12 was found by trial and error and verifying by 
eye that the curve follows the measurements.  
The dependence of the distribution on $ \sqrt{\sigma_\Sigma/\mu_\Sigma}$
is similar to that
expected for a migration distance dependent on the width of the corotation resonance
or if particles approximately move on equipotential curves during the first orbital period 
(e.g., see Figure 6 by \citealt{sellwood02} and \citealt{daniel15}).
The relationship seen here between migration distance and midplane density contrast is 
similar to that found previously between heating rate and spiral Fourier
amplitude by \citet{fujii11}.  

The Fourier spectrograms shown in Figure \ref{fig:fft} exhibit spatial power 
in a range of wavelengths. 
If the spiral structure only contained power at wavelengths significantly larger than the vertical
scale height, the in-plane gravitational potential perturbations would be independent of scale height. 
However we have found that
the dispersion (of $\sigma_{xg}$ at 1 period) depends on the vertical scale height or 
$\sigma_z/\lambda_{crit}$.  We attribute this dependence to the presence of  spatial
power in small scale spiral structure  that is comparable in wavelength to 
the disc thickness.

In the next section, we will extend
the function describing the standard deviation
$\sigma_{xg}$ at $t=1$ and given in equation \ref{eqn:yy} to depend on time 
to explore how $\sigma_{xg}$ depends on time.
  
\subsection{The time dependence of the guiding centre distribution widths}
\label{sec:time}

\begin{figure}
\includegraphics[width=3.5in]{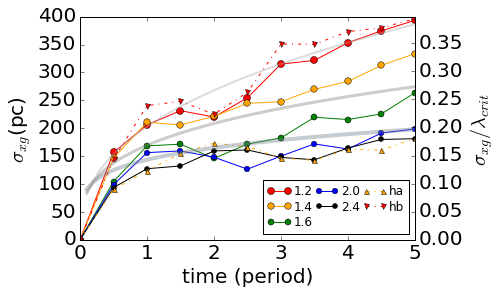}
\caption{
We show the standard deviation $\sigma_{xg}/\lambda_{crit}$ of
 guiding radii for tracer particles as a function of time for simulations
X1--X5, X3hb, and X3hb.
The grey lines show  functions that depend on a power of time (equation \ref{eqn:beta})
with exponents $0.4,0.3,0.2$ for each line, from top to bottom.
The solid points and coloured lines for the X1--X5 simulations are identical to those
previously shown in Figure \ref{fig:sig_xg}.
The $y$ axis is shown in units of pc for a mean surface density
$\Sigma_0 =10 M_\odot {\rm pc}^{-2}$  (left axis) and in units of Toomre wavelength
$\lambda_{crit}$ (right axis).
\label{fig:sig_xg_model}}
\end{figure}

In Figure \ref{fig:sig_xg_model} we plot standard deviations of guiding radii
as a function of time for simulations
X1--X5, X3hb, and X3hb compared with power law curves.
The curves are described by the function 
\begin{equation}
g (\delta\! \rho, t) =  0.12 (\delta\! \rho)^\frac{1}{2} t^\beta \label{eqn:beta}
\end{equation}
where $t$ is in orbital periods and we have extended equation \ref{eqn:yy}
to depend on time.
Three grey lines are shown in Figure \ref{fig:sig_xg_model}. 
The topmost grey line shows the function in equation \ref{eqn:beta} 
evaluated using $\delta\! \rho $ computed from the X5 simulation
at 1 orbit
and a power $\beta = 0.4$, and it is near the red points showing the X5 simulation.  
The middle grey
line uses $\delta\! \rho $ computed from the X3 simulation but exponent $\beta = 0.3$
and it is near the green points showing the X3 simulation.
The lower grey line 
has $\delta\! \rho $ computed from the X1 simulation but $\beta = 0.2$
and it is near the black points showing the X1 simulation.
The X3ha, X3hb simulations are consistent with being
near the bottom and top grey lines, as expected from their density contrasts.
The values of $\beta$ for the grey lines in Figure \ref{fig:sig_xg_model} 
were found by plotting different values
of $\beta$ and  determining by eye if they were near the measurements.

The time dependent behavior seen in 
the width of the guiding centre distribution 
$\sigma_{xg}(t)$ suggests that the exponent $\beta$ is higher when the spiral density contrast is higher.
However further work is needed to verify this as we suspect the lower Toomre $Q$-parameter simulations
 have more rapid changes in spiral morphology during the simulation.  A trend in the value of the
 exponent may be due to time dependent variations in the spiral density contrast as a function of time
 rather than how the migration rate depends on the spiral density contrast itself.

Despite its uncertainty, the exponent $\beta$ appears to be robustly less than 1/2, the 
expected exponent
for a diffusive process giving a random walk in $x_g$.  One possible cause for
this is a reduction in migration efficiency at higher epicyclic amplitude \citep{daniel15}.
Diffusive models for heating account for shallow exponents in this way \citep{carlberg85,jenkins90}.
Though the dependence on epicyclic amplitude does not  explain the difference in the power law 
exponents for the different simulations, it could account for
a reduction in the values of the exponents themselves.

\begin{figure}
\includegraphics[width=3.5in]{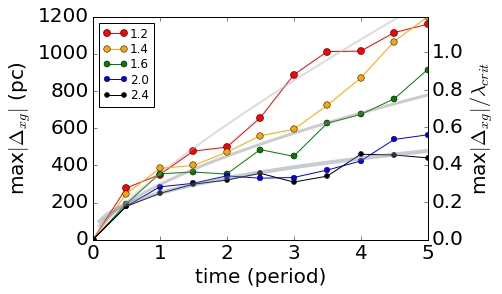}
\caption{
We show  the maximum absolute value of the change in the 
$x$ component of the guiding radius  for tracer particles as a 
function of time for the simulations X1--X5.
The grey lines show  functions that depend on a power of time (equation \ref{eqn:beta2})
with exponent $0.8,0.6,0.4$ for each line from the top to bottom one.
The exponents used are twice those of the curves shown in Figure \ref{fig:sig_xg_model}.
Colours for the lines and points are the same as in Figure \ref{fig:sig_xg} and \ref{fig:sig_xg_model}.  
The $y$ axis is shown in units of pc for a mean surface density
$\Sigma_0 =10 M_\odot {\rm pc}^{-2}$  (left axis) and in units of Toomre wavelength
$\lambda_{crit}$ (right axis).
\label{fig:max_xg}}
\end{figure}

The migration standard deviations are higher for the X5 simulation than the X1 simulation by
about a factor of 2.
To estimate actual distances migrated, we multiply unitless values
 shown on the right hand side of Figures \ref{fig:sig_xg_model}, \ref{fig:max_xg} 
 by the Toomre wavelength.
As a consequence the migration distances are more sensitive to the mean surface density (through
the dependence of the Toomre wavelength)
than they are to the amplitude in spiral structure.
As $\lambda_{crit} \propto \Sigma/\kappa^2 $ and $\Sigma$ is likely to
be exponentially dropping with increasing radius, migrations distances are likely to 
be further in the inner galaxy than the outer galaxy.

The dependence on a critical wavelength was previously proposed and used by \citet{schonrich09}
to describe `churning' with a stochastic model (see their section 2.5 just above their equation 7).
While we confirm a postulated strong dependence on the Toomre wavelength (often called the
critical wavelength and defined in equation \ref{eqn:lambda_crit}), 
\citet{schonrich09}
used a wavelength $\lambda \equiv \sigma_{vR} Q/\kappa = \sigma_{vR}^2/(\pi G \Sigma)$
which differs from the Toomre wavelength.
Perhaps there is a typographic error in their definition and they meant $\lambda = \sigma_{vR}/(Q\kappa)$.
This when multiplied by a factor of $2\pi$  would be equivalent to 
the Toomre wavelength. 

\subsection{Maximal migration rates and distances}
\label{sec:max}

Above we have looked at the time dependence of the width of the distributions
of the $x$ component of the guiding centres.
This ignores the tails of the distribution.    In  section  \ref{sec:oc} below we discuss 
a few open clusters discovered  to be both young and have super-solar metallicity.
These clusters are outliers, with metallicity above most other open clusters.
They could have been born interior to the Sun and migrated outward.
Because they are outliers,
super-solar metallicity young open clusters may have experienced 
more rapid migration than other clusters and so might be in the tail of the distribution
of migration distances, and 
 in  a class dubbed `extreme migrators' by \citet{grand12}.    
In Figure \ref{fig:max_xg} we show the maximum of the absolute value of the change in
 $x_g$ as a function of time for tracer particles in the X1--X5 simulations. 

A distribution of random walkers has standard deviation that grows with $\sqrt{N}$
where $N$ is the number of steps, but
the maximum distance that a walker can travel is proportional to $N$.
The maximum distance is the square of the standard deviation and this remains true
as a function of time.
The steeper slopes  seen in the extreme values shown in Figure \ref{fig:max_xg} compared 
to the standard deviations in Figure \ref{fig:sig_xg} could be due to this scaling.
The grey lines on Figure \ref{fig:max_xg} are curves given by a function 
similar to that  describing the standard deviations, equation \ref{eqn:beta}),
\begin{equation}
h (\delta\! \rho, t) =  0.21 (\delta\! \rho)^\frac{1}{2} t^{2\beta} . \label{eqn:beta2}
\end{equation}
The exponents for each line in Figure \ref{fig:max_xg} are twice those used 
in Figure \ref{fig:sig_xg_model}.  The comparative time dependent behavior  of standard deviation
and maximum are similar to that expected from diffusive-like behavior.
Since the exponents for time are double for the maximum
than for the standard deviation of the $x_g$ distribution,
we tried using a function $\propto \delta\! \rho$ instead of 
$\propto (\delta\! \rho)^\frac{1}{2}$.   However a function $\propto \delta\! \rho$ did not match the
numerical measurements as well as one   $\propto (\delta\! \rho)^\frac{1}{2}$.  
Perhaps the process that sets the guiding centre distribution at 1 orbit
differs from the subsequent diffusive behavior.

In an axisymmetric  disc galaxy, a star in 
a circular orbit has a constant
vertical component of angular momentum as it rotates around the galaxy.
A spiral arm gives a non-axisymmetric (dependent on the azimuthal angle) 
and time dependent perturbation on the star. 
Instantaneously the torque on the star caused by the spiral pattern is
$$\dot L = \frac{\partial V_s} {\partial \theta}$$
where $V_s$ is the perturbation to the gravitational potential caused by the pattern,
and $L$ is the vertical component of angular momentum.
As $L \sim r v_c$ for $v_c$ the velocity of rotation and $r$ the guiding radius,
the torque corresponds to a migration rate in radius $\dot r \sim \dot L/v_c$.
A maximal migration rate can be estimated from this torque 
assuming that the star stays near and on one side of the spiral arm, 
either leading or
lagging the arm, and as shown to be
true for the rapid migrators identified in simulations by \citet{grand12,comparetta12}.
Using a Gaussian bar model for a spiral arm,
\citet{comparetta12} estimated that a linear density enhancement in the disc surface density
with peak density $\Sigma_p$  and oriented with a pitch angle $\gamma$
could cause a maximal migration rate
\begin{align}
\dot r_{max} &\sim  G( \Sigma_p- \mu_\Sigma) 2 \sqrt{\pi} \Omega^{-1} \sin \gamma  \\
& = \lambda_{crit} \Omega  \, \left( \frac{\Sigma_p- \mu_\Sigma} {\mu_\Sigma} \right)
\frac{\kappa^2}{\Omega^2} \frac{\sin \gamma}{2 \pi^\frac{3}{2}} \\
&\approx 2\ {\rm kpc\ Gyr}^{-1} \left( \frac{ \Sigma_p -\mu_\Sigma}{10 M_\odot {\rm pc}^{-2}} \right)
\left(\frac{\sin \gamma}{\sin 24^\circ} \right) \nonumber \\
& \qquad \times
\left( \frac{  \Omega}{ 30\ {\rm km\ s}^{-1} {\rm kpc}^{-1}} \right)^{-1} \label{eqn:rmaxv}
\end{align}
(see their section 4.3). 
Here the pitch angle $\gamma $ is the angle between the ridge of peak density 
and the direction of rotation.  This angle is the same as that
used to characterize spiral arms and the angle is small for a spiral arm that is tightly wound. 

We can replace $\Sigma_p - \mu_\Sigma$ in equation \ref{eqn:rmaxv}
with the standard deviation of the surface density distribution
 $\sigma_\Sigma$ measured in our simulations and write the maximum migration rate 
in units of critical wavelength per rotation period, $\lambda_{crit}/P$
\begin{align}
\frac{\dot r_{max}}{\lambda_{crit} P^{-1}} \approx \sin \gamma \frac{\sigma_\Sigma}{\mu_\Sigma}. 
\end{align}
In our shearing sheet
$\gamma$ is the angle between a linear spiral feature's ridge of peak density and the $y$ axis.
The tilt angle of spiral structures in the shearing sheet simulations are  $\approx 30^\circ$
so for our shearing sheet simulations $\sin \gamma \sim 1/2$.
With surface density contrast $\frac{\sigma_\Sigma}{\mu_\Sigma}$ ranging from 0.2
to 0.5 (listed in Table \ref{tab:simlist}), the maximum migration rates we expect
in our simulations are $\sim 0.1$  to $ 0.25 \lambda_{crit}$ per period, giving
a distance of 0.5 to 1.25  $\lambda_{crit}$ in 5 periods.
Even though the time dependence of the maximum distance migrated
is not linear (the curves in Figure \ref{fig:max_xg} range from $\propto t^{ 0.4}$  to $t^{0.8}$),
the Gaussian bar model estimate for the {\it maximum} migration rate is similar
to those measured in our simulations.  

To achieve a maximal migration rate a particle
would have to be continuously leading (or lagging)  spiral features. 
There is a limit on the distance a particle can migrate.
If we simulate  an increasingly larger number of tracer particles we should not see larger
and larger maxima in the  migration distances.
Likewise, increasing the number of stars observed would not necessarily let us find stars
that have migrated larger distances (past the maximum),
though stars perturbed by other mechanisms could be found (such as those ejected
from the galactic centre). 

\FloatBarrier

\section{Application to open clusters}
\label{sec:oc}

As discussed previously there are only weak constraints on the density contrast
in spiral structure in the solar neighbourhood.  Above we have related the maximum
migration distance to two quantities, a surface density contrast arising from spiral structure 
and the Toomre wavelength.   We discuss super-solar metallicity open clusters to attempt to
place constraints on the spiral structure density contrast  in the Galaxy.

The tracer particles in our simulations are point masses.  When applying the results
of our shearing sheet simulations to open clusters, we neglect the size of the open clusters
and assume that their masses are sufficiently small that individually they have not significantly 
perturbed spiral arms.   Cluster dissolution (e.g., \citealt{gieles07,martinezbarbosa16}) is neglected.

A compilation from the literature 
of age, orbit and [Fe/H]  for super-solar metallicity open clusters that are younger
than 1 Gyr old is listed in Table \ref{tab:tab1}.
We use the notation in brackets to indicate abundances relative to the Sun, i.e., 
$[X/Y] = \log (X/Y) - \log (X/Y)_\odot$ and we use iron, or [Fe/H], to characterize metallicity.
We list apocentre and pericentre radii ($R_a, R_p$) for each cluster computed by \citet{gozha12}.
With an epicylic approximation, the $z$ component of the angular momentum
is set by the guiding radius $R_g$, which is approximately the midpoint radius;  $R_g \approx (R_a+R_p)/2$. 
The difference between current guiding radius and birth guiding radius would be a migration distance.

\begin{table}
\vbox to80mm{\vfil
\caption{\large  Super solar  open clusters younger than 1 Gyr \label{tab:tab1}}
\begin{tabular}{@{}lllllll}
\hline
cluster   &  $R_a$ & $R_p$ & $e$ & $z_{max}$ & age & [Fe/H]  \\
              & kpc       &    kpc    &       & kpc          & Gyr   &  dex  \\
\hline   
NGC 6583 & 6.6   &  5.4   &    0.09  & 0.13  & 1$^a$    & 0.4$^b$  \\
Berkeley 81 &  5.9  & 4.9   &  0.09  & 0.19 &    0.9$^c$ & 0.23$^c$ \\
NGC 2632 & 8.6   & 6.8   &  0.12   & 0.10   &   0.67$^d$  & 0.16$^d$\\
NGC 6067 &  7.6   &  6.8   & 0.06  &  0.07   &   0.090$^e$  & 0.19$^e$ \\
NGC 2232 & 8.3   & 7.8   &  0.03   &  0.05   &   0.032$^f$   & 0.27$^g$ \\
\hline
\end{tabular}
{\\  Apocentre $R_a$, pericentre $R_p$ radii, orbital eccentricity $e$
and maximum $z_{max}$ of the Galactic orbit are those by \citet{gozha12}
except for Berkeley 81.
For Berkeley 81, $e,z_{max}$ agree with that by \citet{vandeputte10}.
For Berkeley 81 we show the apocentre and pericentre radius 
computed using the mean orbital radius of 5.4 kpc estimated by \citet{magrini15}.
References for ages and metallicities:
$^a$\citep{carraro05};
$^b$\citep{magrini10};
$^c$\citet{magrini15};
$^d$\citet{cummings17};
$^e$\citet{alonsosantiago17};
$^f$\citet{silaj14};
$^g$\citet{monroe10};
}}
\end{table}

\begin{table}
\vbox to80mm{\vfil
\caption{\large  Maximal migration rates \label{tab:tab2}}
\begin{tabular}{@{}llllllll}
\hline
cluster        &  $R_g$  & $R_{birth}$ & $d_{mig}$ & age & migration rate  \\
                  &  kpc   & kpc    & kpc & orbits & kpc/Gyr    \\
     \hline
NGC 6583 & 6.0   &  2.5   &   3.5  & 6.5 &3.5   \\
Berkeley 81 &  5.4  &  4.9   & 0.5 & 6.5 &0.5  \\
NGC 2632 & 7.7   &  5.9  & 1.8  & 3.4  &2.7   \\
NGC 6067 &  7.2   & 5.4  & 1.7  & 0.5 & 19   \\   
NGC 2232 & 8.0   &  4.3  &   3.6  &  0.16 & 120 \\ 
\hline  
\end{tabular}
{\\ The  mean or guiding radius is estimated  as $(R_a+R_p)/2$.
Birth radius, $R_{birth}$, is estimated using the [Fe/H] listed in Table \ref{tab:tab1}, $R_g$
and the metallicity gradient -0.07 dex/kpc \citep{anders17}.
The migration distance $d_{mig}$ is  estimated as $R_g - R_{birth}$. 
The migration rate is the migration distance divided by the cluster age.
The cluster age is given in orbital periods at $R_g$ assuming a flat rotation
curve and an orbital period of 210 Myr at $R_g = 8.2$ kpc.
}}
\end{table}

Using the metallicity gradient in [Fe/H] of $\approx -0.07$  dex/kpc  for stars younger
than 4 Gyr by \citet{anders17}
to estimate open cluster birth radii.   This gradient is
based on low galactic latitude red giants with astero-seismic 
estimated ages.
See section 5 and Figure 5 by \citealt{anders17} for a discussion on the sensitivity of the gradient
to age and a comparison of their estimated gradient to those of
other stellar populations, including Cepheids.
Taking  a mean value of [Fe/H] $\approx 0.0$ near $R_\odot$, the gradient
 -0.07  dex/kpc and the metallicity
listed in Table \ref{tab:tab1}, we estimate that NGC 6583 would have been born at a galactocentric radius
of 2.5 kpc.   Taking its current mean galactocentric radius as its guiding or mean orbital radius 
(the average of apocentre and pericentre radii listed in Table \ref{tab:tab1})
we estimate that the cluster could have radially migrated 3.5 kpc.  Using its age this gives
a roughly estimated  migration rate of 3.5 kpc/Gyr.   Similar estimates
for the maximal migration rates are listed in Table \ref{tab:tab2} for the clusters compiled in 
Table \ref{tab:tab1}.

If the metallicities and ages for the two youngest open
clusters, NGC 6067 and NGC 2232 are robust then the needed migration rate is so
high that migration alone cannot account for their super-solar metallicities. 
As a consequence we stop discussing these two clusters in the context of migration.
A local (or patchy) enrichment model (e.g., \citealt{malinie93}) might be explored to account for them.

For the other three older open clusters a migration rate of 0.5 to 3 kpc/Gyr might be required for them
to be born in a higher more metal rich galactocentric radius consistent with their metallicities and 
subsequently migrate outward to their current guiding radii.
NGC 2632, also known as the Praesepe cluster, is similar
in metallicity,  age and kinematics to the Hyades cluster.  \citet{pompeia11} speculated
that the Hyades is at apocentre and a 4:1 resonance with a spiral wave increases its eccentricity 
and allowing it to have mean radius 1\.kpc within $R_\odot$ and nearer to its expected
birth radius (based on its super-solar [Fe/H] abundance).  
If the guiding radius used in Table 2 is overestimated for NGC 2632, we would also
have over estimated the maximum migration extent and rate.  Berkeley 81 has guiding radius
fairly near its expected birth radius so significant radial migration is not needed to account
for its metallicity.  NGC 6583 has a  metallicity high enough to place its 
estimated birth radius within the Galactic bar.  Either it was born within the bar and
the bar  helped eject it from the inner Galaxy, or it was born near the bar end and the
extent of migration required is similar to that estimated above or 2-3 kpc/Gyr. 
NGC 6583 has such a high metallicity it must have migrated outward.   
The metallicities of NGC 2632 and the Hyades suggest that they might have been born
at smaller galactic radii, 1 to 2 kpc smaller than their current guiding radii.

Below we use our simulations to determine what type of spiral structure can
induce the migration distances and rates estimated for the three open clusters NGC 6583,
NGC 2632 and the Hyades (with similar kinematics to NGC 2632).

Our simulation figures show distance in units of the Toomre wavelength and in
pc for a mean surface density of $\Sigma_0 = 10\ M_\odot {\rm pc}^{-2}$ (corresponding
to a Toomre wavelength of $\lambda_{crit,0}  =1007$ pc). 
To estimate standard deviations
in radial migration distances and maximal migration distances we must multiply distances
from our Figures
in units of the Toomre wavelength by the Toomre wavelength for the disc
that is causing the migration.   Alternatively
we can use the distances in pc if we multiply them by the ratio of surface densities
(the ratio of the disc that is causing the migration to $\Sigma_0 = 10 M_\odot {\rm pc}^{-2}$
or the ratio of the Toomre wavelength of the disc causing the migration to
$\lambda_{crit,0} = 1007$ pc).
Curves fitting both standard deviations and maximal migration distances also depend on
the square root of the spiral density contrast (as in equations \ref{eqn:beta} and \ref{eqn:beta2}).
At a fixed surface density (mass per unit area), and using an exponential vertical density profile, 
the density (mass per unit volume) in the midplane is inversely proportional to the vertical scale height.
Thickness of galactic components are often given in terms of a scale height  $h$ assuming
the density distribution is proportional to ${\rm sech}^2(z/h)$ (expected for an isothermal, self-gravitating disk)
or $\exp (-|z|/h) $  (expected for isothermal stars or gas in a constant gravitational field).
The density profile of our simulations does not have a sharp peak at $z=0$ and is self-gravitating
so we estimate the scale height of our simulations from the standard deviation of the sech$^2$ function,
giving standard deviation $\sigma_z \approx 0.9 h$.  So the scale heights in our simulations
can be estimated using values for $\sigma_z$ in  Table \ref{tab:common} with $h \approx 1.1 \sigma_z$.
We can correct for a difference in disc thickness, the Galactic disc as compared
to that of our simulations,  with a factor that depends on the square root
of the ratio of scale heights (that of the disc causing the migration and that used
in the simulation).
Thus we can predict migration distances using our figures if we multiply them by a factor
that depends on the ratio of the Toomre wavelength (compared
to 1007 pc)  times  the square root of the ratio of scale heights.

We contrast the role of large amplitude variations in a lower mean surface density  gas disc
with the role of smaller amplitude variations in a higher surface density stellar disc.
We start by predicting migration distances caused by a gas disc.
Our vertical standard deviation for our simulations was $\sigma_z \approx 150$ pc 
(see Table \ref{tab:common})
corresponding to  a scale height of about  of $h \sim165$ pc.  
This exceeds a gas scale height 
(for molecular and cold atomic hydrogen) in the solar neighbourhood
of about 100 pc \citep{mckee15}. 
The surface density in molecular and cold atomic hydrogen in the solar neighbourhood 
is somewhat lower
than $\Sigma_0 = 10 M_\odot {\rm pc}^{-2}$, or about $\Sigma = 7 M_\odot {\rm pc}^{-2}$  \citep{mckee15}.  
We can use distances in our figures in units of pc if we correct these distances
by the ratio of $\Sigma/\Sigma_0 = 7/10$ (for the Toomre wavelength) and by the ratio of $\sqrt{165/100}$
to take into account the difference between the gas scale height and
that of the simulations.
The two corrections to the X1--X5  simulations
approximately cancel each other out ($7/10 \times \sqrt{165/100} \approx 0.9$), 
so we can use the distances in pc
in our figures directly for comparison to open cluster estimated migration distances.
 As we expect large
variations in gas density due to spiral structure 
we choose the simulation with highest density contrast or the X5 simulation for comparison.

 After 3 orbital periods the standard deviation in $x_g$ in the X5 simulation
would be only about 300 pc  and at 5 periods, 400pc (from inspection of Figure \ref{fig:sig_xg_model}.  
The most extreme outliers have a maximum
migration distance of about 900 pc at 3 periods and 1.2 kpc at 5 periods
 (using Figure \ref{fig:max_xg}).
These migration distances are not larger enough to account for the 
estimated needed distances for migration for NGC 2632 (Praesepe) and Hyades
clusters  (we estimated 1.8 kpc over 700 Myr or 3.4 orbital periods).  Likewise we fall short
for NGC 6583 (needing 3.5 kpc at 1 Gyr but in the inner galaxy, at 6 orbital periods).
We conclude that spiral structure in the gas disc {\it alone} cannot induce sufficient
migration to account for young super-solar metallicity open clusters.

The stellar surface density is higher than the gas surface density. 
Taking a value
for the stellar surface density of about
$\Sigma \approx 33 M_\odot {\rm pc}^{-2}$ (taking the value for $\Sigma_*$ from Table 3 by \citealt{mckee15}), 
the Toomre wavelength is $\lambda_{crit} \approx 3.3$ kpc, 
exceeding by a factor of about 3 the value we used to give distances in units of pc in our figures.
At this Toomre wavelength the standard deviation in the $z$ density distribution in the X series simulations is
$\sigma_z = 500$ pc (multiplying the value from Table \ref{tab:common} by the Toomre wavelength)
or a vertical scale height of $h \sim 550$ pc.
The stellar scale height is about 400 pc in the solar neighbourhood (taking values for M dwarfs by \citealt{mckee15}).
We should correct distances in our simulations by the square root of the scale
height ratio or 1.17.   
We should correct distances in pc in our figures
by the square root of the scale height ratio or 1.17 and by  $\Sigma/\Sigma_0= 3.3$ 
to take into account
the Toomre wavelength, or a factor of about $1.17 \times 3.3 \sim 3.8$.
 
\begin{figure}
\includegraphics[width=3.4in]{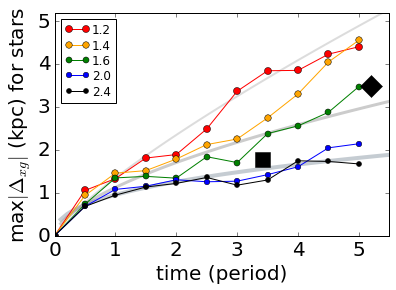}
\caption{
We show the maximum absolute value of the change in the 
$x$ component of the guiding radius  for tracer particles as a 
function of time for the simulations X1--X5, but now scaled to
a stellar disc of $\Sigma = 30 M_\odot {\rm pc}^{-2}$ and with
vertical scale height similar to the Galactic thin disc.
The black square shows estimated migration distances
for NCC  2632 (Praesepe) and Hyades clusters.  The black diamond shows
estimated migration distance for NGC 6583.  The black square
is placed at the age of the NCC  2632 and Hyades clusters in orbits.
The black diamond should be at 6.5 orbits but this lies off our integrations so
we have placed it at 5.2 orbits.
The grey lines and circles are from the X1-X5 simulations and the same
as in Figure \ref{fig:max_xg}.
This figure shows that a simulation with spiral density contrast similar
to the X3 simulation has maximal migration distances similar to the super-solar
metallicity  Praesepe, Hyades and NGC 6583 open clusters.
\label{fig:max_oc}}
\end{figure}

\begin{figure}
\includegraphics[width=3.5in]{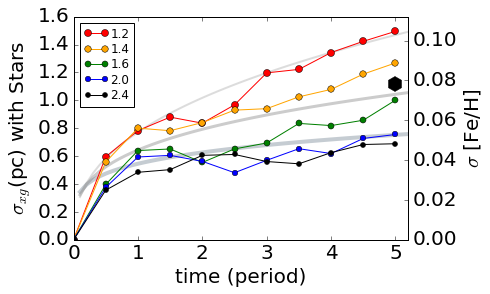}
\caption{
We show the standard deviation $\sigma_{xg}$ of
 guiding radii for tracer particles as a function of time for simulations
X1--X5, but now scaled to
a stellar disc of $\Sigma = 30 M_\odot {\rm pc}^{-2}$ and with
vertical scale height similar to the Galactic thin disc.
The grey lines and circles are from the X1-X5 simulations and the same
as in Figure \ref{fig:max_xg}.
The right axis has converted units of distance to variation  in [Fe/H]
using the metallicity gradient by \citealt{anders17} of -0.07 dex/kpc.
The black hexagon shows the standard deviation in metallicities for 38 open clusters
with  high-quality spectroscopic data that were compiled by \citet{netopil16}.
This figure shows that a simulation with spiral density contrast similar
to the X3 simulation predicts a standard deviation in migration distance consistent with the metallicity
scatter in young open clusters.
\label{fig:sig_FeH}}
\end{figure}

Figure \ref{fig:max_oc} and Figure \ref{fig:sig_FeH} 
 show maximal migration distances and standard deviations in guiding centre
for the X1-X5 simulations rescaled by the factor 3.9.
The spiral amplitudes in stars would
be lower than for the gas, perhaps similar to the X3 simulation.  We will discuss 
this choice in the next three paragraphs. 
Using the X3 simulation
we take values in pc at 3 orbits and 5 orbits from Figure \ref{fig:max_oc} and \ref{fig:sig_FeH}.
giving  standard deviations $\sigma_{xg} =$ 0.8  and 1.0 kpc, and  
the maximum migration distances  2.0  and 3.1 kpc.
These values exceed those estimated for the gas disc.
Even though the gas density might have larger density variations,
we estimate that  low amplitude spiral structure in the more massive stellar disc causes more
radial migration.  
 
With ages corresponding to 3 orbital periods, 
the Hyades and Praesepe clusters  require about 1.8 kpc of
migration from their birth radii,
and this is similar to the maximum migration distance seen in the X3 simulation. 
To illustrate this we have placed a black square onto Figure \ref{fig:max_oc}
to represent these two clusters.
The standard deviation of the guiding centre distribution estimated from our simulations
(0.8 kpc at 3 orbital periods) is below that
required for these two clusters, but this would be consistent with the rarity of super-solar metallicity open
clusters in the solar neighbourhood (as the standard deviation must be lower than the 
absolute value of the maximum migration distance).
At 5 orbital periods the maximum distance reached (using the X3 simulation) is 3.1 kpc
and this is similar to that required to account for NGC 6583, which is shown
as a black diamond on Figure \ref{fig:max_oc}.  The standard deviation
at 5 orbital periods is about 1\. kpc.  So at 1 Gyr most clusters would lie within 1 kpc
of their birth radius, putting NGC 6583 in the tail of the distribution. 
Were we to use
a higher mean surface density, appropriate in the inner galaxy, our estimated maximum migration
distance would be even larger.  In summary the maximum migration distances 
estimated from the X3 simulation are sufficient to account for migration
distances estimated for these three young and super-solar metallicity open clusters.

Above we estimated that after 1 Gyr, clusters born at the same radius would have a 
 standard deviation in galactic radius due to migration of about 1 kpc.
Using the radial metallicity gradient by \citet{anders17}, this 1 kpc distance corresponds
to a variation in metallicity of about 0.07 dex.   So we would estimate
that the standard deviation of metallicities  of young open clusters would be the same, or about 0.07 dex
in [Fe/H].   
We compare this number to the standard deviation in [Fe/H] of young open clusters.
From the 88 open clusters with high-quality spectroscopic data that were compiled by \citet{netopil16},
38 have mean galactic radius ($R_g$) within 7--9 kpc, and ages less than 1Gyr
(see their Figure 6 showing the radial metallicity distributions).
The standard deviation in [Fe/H] for these 38 clusters is 0.074 dex (where we have used
values for [Fe/H] based on the high-quality spectroscopic data from Table A1 by \citealt{netopil16}).   
Thus our  estimate for the metallicity dispersion induced by migration, using the X3 simulation, is 
consistent with that observed.    Had we used a lower or higher Toomre $Q$-parameter simulations
for comparison (X1, X2 or X4, X5) the predicted standard deviation in radial migration distances would 
not have  agreed with the young open cluster standard deviation in [Fe/H].
The density contrast present in the X3 simulation is consistent both with the dispersion
in [Fe/H] and the few super-solar metallicity young open clusters that represent rarer
more extreme migrators.

Figure \ref{fig:sig_FeH} shows that after 1 orbital period, 
the standard deviation in guiding centre coordinate
$x_g$ does not increase rapidly with time.
Consequently taking into account the open cluster age distribution (for clusters
older than 1 Gyr) would not significantly increase
the estimated metallicity dispersion arising from migration. 
We have not numerically studied migration at times longer than 5 orbital periods,
but the  slow increase in the standard deviation as a function of time seen in our 
simulations would
be consistent with the absence of a strong correlation between open cluster age and metallicity 
(the age-metallicity relation; e.g., \citealt{carraro94,yong12,netopil16}). In other words,
the scatter in [Fe/H] due to migration could exceed the slow increase in metallicity due 
to ongoing large scale enrichment in the disc.
Our estimates for the metallicity scatter 
neglects local (patchy) enrichment of the ISM \citep{balser15,berg15,vogt17,krumholz18}
and this additional process might be required
to account for very young open clusters such as NGC 6067 and NGC 2232 
(and the high [$\alpha$/Fe] cluster NGC 6705; \citealt{magrini14,magrini15,casamiquela17b})
that could not have migrated far enough to account for their abundances.

In summary, the rare super-solar metallicity open clusters near the Sun appear to be consistent with 
a stellar density contrast (for spiral arms) similar to that of the X3 simulation or with a surface density
contrast $\sigma_\Sigma/\mu_\Sigma \sim 0.26^{+0.08}_{-0.04}$.  Here
we have taken the value for $\sigma_\Sigma/\mu_\Sigma$ for the 
X3 simulation (listed in Table \ref{tab:simlist}).  For the uncertainty we have taken
the difference between the density contrast in the X4 and X3 simulations (+0.08)
and between the X2 and X3 simulations (-0.04).
The inferred extent of recent radial migration at this density contrast is also
consistent with the standard deviation in [Fe/H] for young open clusters.
Were we to use a lower value for the metallicity gradient of 0.06 dex/kpc, estimated
migration distances for the clusters would increase by about 20\% and our estimated
density in stellar density contrast in spiral arms would be larger but within
the higher end of the +0.08 uncertainty.

Our spiral arm surface density contrast level $\sigma_\Sigma/\mu_\Sigma \sim 0.26^{+0.08}_{-0.04}$
somewhat exceeds that estimated from
 the Galactic COBE model \citep{drimmel01}
with  peak above mean divided by mean
$(\Sigma_p - \mu_\Sigma)/\mu_\Sigma) \approx 0.16 \pm 0.03$.
Here we have    
taken the number based on the K-band spiral amplitude
discussed at the end of section 6 by \citet{drimmel01} and used the $\pm 18\%$ 
uncertainty listed in their Table 2.
As discussed by \citet{drimmel01},  the COBE model measurement is dependent on the type of model
used to fit the COBE data and  lower than that expected based on
imaging studies of other galaxies, so perhaps we should not be concerned
that our density for the local density contrast is higher than that previously measured.
Also, as we will discuss in section \ref{sec:sum}, there are a number of reasons  
our estimate is not precise.  

In section \ref{sec:max} 
we gave in equation  \ref{eqn:rmaxv} 
an estimate for the maximal migration rate using the Gaussian bar model
\citep{comparetta12}.
We now compute the maximum migration 
migration rate in the Solar neighborhood estimated with this model.
With an  angular rotation rate in the Solar neighborhood
$\Omega_\odot \approx 30 \ {\rm km\ s}^{-1} {\rm kpc}^{-1}$ \citep{bland16}, 
spiral density contrast
$(\Sigma_p- \mu_\Sigma)/\mu_\Sigma \approx 0.16$ based on
the COBE data model \citep{drimmel01}, 
and a mean stellar surface density
$\mu_\Sigma \approx 33 M_\odot {\rm pc}^{-2}$  \citep{mckee15},
the maximum migration rate is $\dot r_{max} \sim 0.5\ {\rm kpc\ Gyr}^{-1}$ for 
 a pitch angle of $12^\circ$  and about $1\ {\rm kpc\ Gyr}^{-1}$ for a pitch angle
 of $24^\circ$, spanning the pitch angle estimates compiled by  \citet{vallee08}.
These maximal migration rates are lower than the required open cluster
migration rates listed in Table \ref{tab:tab2}
for NGC 6583, and the Praesepe and Hyades open clusters.  This rough estimate
for the maximal migration rate
supports our inference that the spiral density contrast in the solar neighborhood 
could be higher than that estimated by the COBE model.

\section{Discussion and Summary}
\label{sec:sum}

We use shearing sheet N-body simulations to investigate how far
stars and open clusters can migrate in a galaxy disc within 5 orbital periods. 
The simulations contain massive particles that exhibit spiral structure due to their
own self-gravity.
Massless tracer particles are inserted into the simulation after spiral structure has grown.
Due to perturbations from the spiral structures, guiding centres of
the tracer particles drift.  This drift corresponds to radial migration in a disc galaxy.
 As a function of time, we measure the width and maximum of the distribution of the changes in
guiding centres  for tracer particles.
 
The shearing sheet simulations suggest that the rate and extent of radial migration
is primarily set by the Toomre or critical wavelength.
As this wavelength is set by  the mean surface mass density,
 migration rates and extent should be higher in the inner regions of galaxies
than the outer regions.     We find that in 5 rotation periods, the standard deviations
of the guiding centre distributions broaden to between 0.2 and 0.4 of the Toomre wavelength
with a maximum migration distance (in guiding centre $x$ component) about three times this. 

To a lesser extent migration rates depend on
the surface density or midplane density contrast in the spiral structure.
The standard deviations of the guiding centre distributions
can be described by power laws with exponents in the range 0.2 to 0.4.
The maximal distances obey exponents twice those
of the standard deviation suggesting that a diffusive model may describe the  behavior of the 
guiding centre distributions.  A diffusive model could operate on such a short time
 if individual spiral features are uncorrelated or if patterns interfere with one another. 
We attribute
the variation in the exponents in different simulations to slow variations in the spiral morphology 
 that was more rapid in our lower Toomre $Q$-parameter simulations.

We used our simulations and estimated guiding radii for super-solar metallicity open clusters 
to attempt to constrain the surface density contrast
in spiral structure at the Sun's galactocentric radius.
The comparison suggests that the surface density contrast 
has ratio of standard deviation to mean $\sigma_\Sigma/\mu_\Sigma \sim 1/4$
with an uncertainty of about 30\%, and with value
somewhat exceeding the COBE model by \citet{drimmel01}.

Our estimate for the density contrast is uncertain for a number of reasons.
Spiral structure in our simulations remains near corotation.  
 In contrast, galactic simulations
of a disc often show patterns that move with different frequencies (e.g., \citealt{quillen11}).
Spiral spiral patterns can influence  each other or be coupled to a galactic bar.
A bar or spiral pattern that is distant from its corotation might still affect radial
migration \citep{brunetti11,minchev12}.   
 Nearby features could interfere causing peaks to drift in radius 
or increase and decrease in amplitude (e.g., as described by \citealt{quillen11}).
It is difficult to say whether these effects would increase or decrease the extent of
radial migration compared to that predicted using the shearing sheet.   Perhaps a study
similar to \citet{fujii11} that focused on heating could improve upon estimates for
migration using complete disc simulations.

The expression for the Toomre wavelength (equation \ref{eqn:lambda_crit}) 
implies that  near the Sun's galactic radius,   $R_\odot$, the Toomre
wavelength is large,   $\lambda_{crit} \sim R_\odot$.  
The shearing sheet approximates a local patch of a rotating disc (in polar coordinates)
with a Cartesian square (see Figure \ref{fig:shearsheet}) and neglects radial gradients
in epicyclic frequency, mass surface density and velocity dispersion.  
While shearing sheet simulations are internally consistent (in that their spiral 
structure is evolving due to their own self-gravity)
they may not be a good approximation for applications, such as discussed here, requiring
 a  shear box size similar to or larger than the associated galactic radius or the disc scale length.
Full disc simulations are required to improve upon applications derived from  
shearing sheet simulations.
The Toomre wavelength, though it is a locally computed quantity, may
also be  relevant to larger scale spiral morphology.
 \citet{donghia15} proposed that the Toomre wavelength locally sets
the number of spiral arms, hence the large  Toomre wavelength computed at and within the
Sun's galactic  radius
could account  for the small number  of spiral arms (2 to 4) within the Sun's galactic radius.

Our simulations only contain point masses.  A  gas and stellar disc could behave differently
than the phenomena we see in pure N-body simulations.
Our tracer particles were inserted abruptly into a simulation containing spiral structure
with an ad hoc damping to help reduce Toomre $Q$-parameter variations in the disc.
Their initial velocity is not a good match to those of recently born stars.
The Galaxy is comprised of multiple populations, each with a different scale height.
We have neglected this structure, relating the migration rate only to a vertical dispersion
and a spiral density contrast.
Future numerical studies could reduce these errors and uncertainties with more detailed simulations.
While open clusters facilitate measurement of both metallicities and ages, their distributions
may be biased if cluster number and evaporation and destruction 
are correlated with age, birth radius or metallicity.
Uncertainties in cluster orbit, metallicity, the extent of local ISM enrichment or abundance
variations, and the metallicity gradient also affects estimated migration distances.
Better numerical simulations would allow 
more detailed observations and improved measurements 
of young stars and open clusters  
to be placed quantitatively in context with improved models for migration.

\vskip 2 truein  
ACQ is grateful to the Leibniz-Institut f\"ur Astrophysik Potsdam for their 
warm welcome and hospitality July 2017, Mt. Stromlo observatory
for their warm welcome and hospitality Nov 2017-- Jan 2018, 
and support from the Simons Foundation.
This material is based upon work supported in part by the National Science Foundation
Grant No. PHY-1460352,
for the University of Rochester's Department of Physics and Astronomy REU program.
C.C. acknowledges support from DFG Grant CH1188/2-1 and from the ChETEC COST Action
(CA16117), supported by COST (European Cooperation in
Science and Technology).
We thank Marica Valentini, Friedrich Anders, Charlie Lineweaver, Agris Kalnajs, Mark Krumholz,
and David Hogg for helpful discussions.  
We are grateful to Agris Kalnajs for making his conference proceeding available to us
electronically and for re-examining his shearing sheet simulations from 1991.

\appendix
\section{The shearing sheet approximation}
\label{sec:app1}

The Hamiltonian for the shearing sheet is
\begin{align}
H(p_x, p_y; x, y) &= \frac{p_x^2}{2} + \frac{p_y^2}{2} - 2 p_y \Omega x  + \frac{ \kappa^2 x^2 }{2}.
\label{eqn:Ham}
\end{align}
The momenta $p_x,p_y$ are canonical and conjugate to the coordinates $x,y$.
The epicyclic frequency $\kappa$ determines the frequency of  oscillations in the $x$ direction.
This Hamiltonian can be derived by writing a 2D Hamiltonian for particle motion 
in cylindrical coordinates for motion in the plane and transforming to a frame rotating with the disc
at a particular radius.  A patch is chosen centered at this radius.    The Hamiltonian is then 
expanded to second order in coordinates from the center of the disc patch and to second
order in canonical momenta.
In celestial mechanics
this is known as Hill's approximation (e.g., \citealt{rein12}) and is equivalent
to a classic epicyclic approximation (e.g., \citealt{B+T}).

This Hamiltonian gives equations of motion 
\begin{align}
x &= x_g  +  C \cos (\kappa t + \phi_0) \nonumber  \\
v_x &= -  C \kappa \sin (\kappa t + \phi_0) \nonumber  \\
y &= c - \frac{(4 \Omega^2 - \kappa^2)}{2\Omega} x_g t - \frac{2 \Omega}{\kappa}  C \sin (\kappa t + \phi_0) \nonumber  \\
v_y &= - \frac{(4 \Omega^2 - \kappa^2)}{2\Omega} x_g - {2 \Omega} C \cos (\kappa t + \phi_0).
\label{eqn:motion}
\end{align}
Here particle velocities $v_x = \dot x, v_y  = \dot y$.
The orbit can be described in terms of a guiding centre, $x_g,y_g$ and epicyclic oscillations $x_s,y_s$
about the guiding centre $x  = x_g  + x_s; y=y_g + y_s$.
The parameter $C$ can be recognized as a constant epicyclic amplitude 
and the parameter $\phi_0$ as a phase (see for example
section 3.2.3 by \citealt{B+T}).  
The  guiding centre $x_g$ position is also a constant.
The $y_g$ guiding centre coordinate drifts depending on the velocity shear,
\begin{equation}
y_g = c - \frac{(4 \Omega^2 - \kappa^2)}{2\Omega} x_g t \label{eqn:shear}
\end{equation}
where the parameter $c$ is a constant.
With zero epicyclic amplitude $y = c  - \frac{(4 \Omega^2 - \kappa^2)}{2\Omega} x t$,
 the velocity is solely in the $y$ direction
$\dot y = - \frac{(4 \Omega^2 - \kappa^2)}{2\Omega} x$  and depends only on $x$ 
(see Figure \ref{fig:shearsheet}).

It is often useful to compute
the guiding centre position in terms of current positions and velocities
\begin{align}
y_g &= y - y_s = y -  \frac{ 2 \Omega}{\kappa^2} v_x \nonumber\\
      &= y -  \frac{ \Omega^2}{\kappa^2} 2\Omega^{-1} v_x  \nonumber\\
x_g &= x - x_s = \frac{4 \Omega^2}{\kappa^2} x  + \frac{2 \Omega}{\kappa^2} v_y\nonumber\\
&= \frac{\Omega^2}{\kappa^2} \left( 4 x + 2 \Omega^{-1} v_y \right). \label{eqn:xg}
\end{align}
The epicyclic amplitude $C$ and phase $\phi$ can then be computed using the guiding $x_g$ and positions and
velocities with
\begin{align}
C^2 &= (x - x_g)^2 + v_x^2/\kappa^2  \nonumber \\
\phi &= \kappa t + \phi_0 = \texttt{atan2}(-v_x/\kappa, x-x_g)  \label{eqn:Cphi}
\end{align}
where the inverse tangent function is used to compute angles in all quadrants.

\subsection{Modification to the SEI integrator}
\label{sec:app2}

We modify the equations in section 3.5
 by \citealt{rein12} for the SEI integrator so that a non-Keplerian value of the epicyclic frequency $\kappa$
 can be used.
 \citet{rein12} denote
 $x^n,y^n,v_x^n,v_y^n$ for the positions and velocities at timestep $n$.
Their equations  9 for the centre of epicyclic motion
are modified as follows with factors of $\Omega/\kappa$:
\begin{align}
x_0^n &= \frac{\Omega^2}{\kappa^2} \left( 2 v_y^n + 4 x^n  \right) \nonumber \\
y_0^n &= y^n - \frac{\Omega^2}{\kappa^2} 2 v_x^n \Omega^{-1}. \label{eqn:guide}
\end{align}
where $x_0^n,y_0^n$ are the guiding centres at timestep $n$.
The epicyclic vector (their equations 10) is modified to
\begin{align}
x_s^n &= \Omega(x^n - x_0^n)  \nonumber \\
y_s^n &= \frac{1}{2} \frac{\kappa}{\Omega} (y^n - y_0^n).
\end{align}
The epicyclic motion is written as a rotation and their equation 11 is modified to be
\begin{align}
x_s^{n+1} &= x_s^n \cos (\kappa \Delta t) + y_s^n \sin(\kappa \Delta t) \nonumber \\
y_s^{n+1}&= -x_s^n \sin (\kappa \Delta t) + y_s^n \cos(\kappa \Delta t) .  \label{eqn:rotate}
\end{align}
The  guiding centre coordinates are restored by modifying their equations 12 to be 
\begin{align}
x^{n+1} &= x_s^{n+1} \Omega^{-1} + x_0^n \nonumber \\
y^{n+1} &= 2 y_s^{n+1} \frac{\Omega}{\kappa} + y_0^n
    - \frac{1}{2} \left(4 - \frac{\kappa^2 }{\Omega^2} \right) x_0^n \Omega \Delta t\nonumber \\
v_x^{n+1} &= y_s^{n+1} \frac{\kappa}{\Omega}   \nonumber \\
v_y^{n+1} &= - 2 x_s^{n+1} - \frac{1}{2} \left(4 - \frac{\kappa^2}{\Omega^2} \right)   x_0^n \Omega.
\label{eqn:guide_restore}
\end{align}
With $\kappa/\Omega  =1$  for a Keplerian disc equations \ref{eqn:guide} - \ref{eqn:guide_restore}
are the same as the original ones by \citet{rein12}.

The \texttt{rebound} code is modified by adding a new unitless parameter
\texttt{KAPPA\_OMEGA} $ \equiv \kappa/\Omega$.   It is
defined like \texttt{OMEGA} in  \texttt{rebound.h} in the definition
for the structure \texttt{struct reb\_simulation\_integrator\_sei}).
In the routine \texttt{rebound.c} we initialize \texttt{KAPPA\_OMEGA} =1
so there is no change for the user wanting a Keplerian disc.
In routine \texttt{boundary.c} for the cases for the
shear boundary (\texttt{REB\_BOUNDARY\_SHEAR})  occurrences of
$\frac{3}{2} \Omega$ are modified to become
$\frac{1}{2} \left( 4 - \frac{\kappa^2}{\Omega^2} \right)$.
The SEI integrator itself in \texttt{integrator\_sei.c} is modified
 using equations \ref{eqn:guide} - \ref{eqn:guide_restore}.
 
\subsection{Vertical motion}
\label{sec:appz}  
  
The Hamiltonian in equation \ref{eqn:Ham} can be extended to allow motion in the
vertical direction with an additional  momentum $p_z$ and coordinate $z$ and with
an additional term added to the Hamiltonian
\begin{equation}
H_3(p_z,z) = \frac{p_z^2}{2} + \frac{\Omega_z^2 z^2}{2}
\end{equation}
where the velocity $v_z = \dot z = p_z $ and $\Omega_z$ is the vertical epicyclic frequency.
The \texttt{rebound} code allows the vertical epicyclic frequency to be adjusted separately 
from $\Omega$ and $\kappa$
(see equation 13 by \citealt{rein12} and associated discussion).  The equations of  
of motion
\begin{align}
z &= D \cos (\Omega_z t + \varphi_0) \nonumber \\
v_z & = - D \Omega_z \sin(\Omega_z t + \varphi_0) \label{eqn:zvz}
\end{align}
where $\Omega_z$ is the vertical epicyclic frequency, the constant $D$ is the vertical epicyclic amplitude and 
$\varphi_0$ is a phase.  The vertical epicyclic frequency $\Omega_z$ is labelled \texttt{OMEGAZ} in \texttt{rebound}.
The $z$ and $v_z$ coordinates are updated similar to the epicyclic vector in equation \ref{eqn:rotate}
as shown in equation 13 by \citealt{rein12}.
The vertical epicyclic amplitude $D$ and epicyclic phase $\varphi$ can be computed from  
coordinate $z$ and vertical velocity $v_z$
\begin{align}
D^2 &= z^2 + v_z^2/\Omega_z^2 \nonumber \\
\varphi &= \Omega_z t + \varphi_0 = \texttt{atan2} ( - v_z/\Omega_z, z)
\end{align} 
similar to equation \ref{eqn:Cphi}.
 
 \end{document}